\documentclass[12pt]{article}
\pdfoutput=1

\usepackage{jheppub}
\usepackage{amsmath,amssymb,euscript,array,mathrsfs}
\usepackage{slashed}

\usepackage{epsfig}

\def\a{\alpha}
\def\b{\beta}
\def\c{\gamma}
\def\d{\delta}
\def\e{\epsilon}

\def\l{\lambda}
\def\m{\mu}
\def\n{\nu}

\def\s{\sigma}

\def\w{\omega}

\def\D{\Delta}

\def\tr{{\rm tr}}

\def\Dbarslash{\,\,{\raise.15ex\hbox{/}\mkern-12mu {\bar D}}}
\def\Dslash{\,\,{\raise.15ex\hbox{/}\mkern-12mu D}}
\def\delslash{\,\,{\raise.15ex\hbox{/}\mkern-9mu \partial}}
\def\delbarslash{\,\,{\raise.15ex\hbox{/}\mkern-9mu {\bar\partial}}}

\def\half{\frac{1}{2}}
\def\thalf{\tfrac{1}{2}}

\def\rta{\rightarrow}
\def\tr{{\rm tr}}

\title{\begin{center}Lorentz and \textsf{CPT} violation and the hydrogen and antihydrogen
molecular ions \,I -- rovibrational states\end{center}}

\author{\vskip1.3cm
\large{ Graham M.~Shore}}

\emailAdd{g.m.shore@swansea.ac.uk  }

\affiliation{\vskip0.8cm 
Department of Physics, Faculty of Science and Engineering, Swansea University, 
Singleton Park, Swansea, SA2 8PP, UK}

\date{\today}

\abstract{The extremely narrow natural linewidths of rovibrational energy levels in
the molecular hydrogen ion ${\rm H}_2^{\,+}$, and the prospect of synthesising
its antimatter counterpart $\overline{\rm H}_2^{\,-}$, make it a promising 
candidate for high-precision tests of fundamental symmetries such as Lorentz and
\textsf{CPT} invariance.  
In this paper, we present a detailed analysis of the rovibrational spectrum of the
(anti-)hydrogen molecular ion in a low-energy effective theory incorporating 
Lorentz and \textsf{CPT} violation. 
The focus is on the spin-independent couplings in this theory, and especially the
\textsf{CPT} odd couplings for which the
best current bounds come from measurements of the $1S$\,-\,$2S$ transition
in atomic hydrogen and antihydrogen.
We show that in addition to the improvement in these bounds from the increased
precision of the transition frequencies, potentially reaching 1 part
in $10^{17}$, rovibrational transitions in the ${\rm H}_2^{\,+}$ and $\overline{\rm H}_2^{\,-}$
molecular ions have an enhanced sensitivity to Lorentz and
\textsf{CPT} violation of $O(m_p/m_e)$
in the proton (hadron) sector compared to ${\rm H}$ and $\overline{\rm H}$ 
atomic transitions.

\vskip3.5cm}

\notoc

\begin{document}

\maketitle

\setlength{\parskip}{10pt}

\section{Introduction}\label{sect 1}

Local relativistic quantum field theories are the foundation of our current understanding of elementary
particle physics. Together with microcausality, their basic principles of locality and Lorentz invariance
imply the existence of antimatter and the necessity of \textsf{CPT} invariance as an exact symmetry
of nature \cite{Pauli:1940, Bell, Luders, Pauli:1955}.

Given their essential r\^ole in our current theories, it is crucial that these fundamental
principles -- Lorentz invariance, \textsf{CPT} symmetry and locality
-- are tested experimentally to the highest possible precision \cite{Charlton:2020kie}. 
To achieve such ultimate standards of precision, however, it is necessary to look
beyond high-energy particle physics experiments to fundamental atomic physics,
especially spin-precession measurements on elementary particles and atomic spectroscopy.
For example, by comparing cyclotron frequencies, the BASE experiment
at CERN has established the  equality of the charge-to-mass ratios
of the proton and antiproton to 16 parts in $10^{12}$ \cite{BASE:2022yvh},  while the ALPHA 
collaboration has measured the equality of the $1S$\,-\,$2S$ transition in antihydrogen with that 
of hydrogen to 2 parts in $10^{12}$ \cite{Ahmadi:2018eca}, both key tests of \textsf{CPT} invariance.

The progress made in recent years by ALPHA in cooling, trapping and investigating antihydrogen
has opened an era of high precision anti-atom spectroscopy, and future developments will push 
towards the benchmark precision of $O(10^{-15})$ achieved for the $1S$\,-\,$2S$ transition in hydrogen.
Already several transitions have been studied in detail, including $1S$ hyperfine \cite{ALPHA:2017fsh}
and $1S$\,-$2P$ \cite{ALPHA:2020rbx} transitions in addition 
to $1S$\,-\,$2S$ \cite{ALPHA:2016siw, Ahmadi:2018eca}, and the results interpreted theoretically 
in terms of constraints on \textsf{CPT} violation.
% \cite{NaturePhysics}.

In this paper, building on the work of \cite{Muller:2004tc},
we extend the theoretical analysis of Lorentz and \textsf{CPT} symmetry breaking from
(anti-)atoms to (anti-)molecules, in particular the molecular hydrogen ion ${\rm H}_2^{\, +}$ and its
antimatter counterpart, $\overline{\rm H}_2^{\, -}$. The compelling feature of molecular ions for precision
tests of fundamental symmetries is the existence of long-lived, extremely narrow linewidth,
{\it rovibrational} states in which the two bound (anti-)protons make transitions between
energy levels $E_{v N}$ characterised by discrete vibrational and orbital angular momentum states,
labelled here by $v$ and $N$ respectively.\footnote{For a comprehensive recent review see \cite{SchillerCP};
earlier work is described in \cite{Carrington:1989}.}

These rovibrational transitions offer the possibility in principle
of testing Lorentz and \textsf{CPT} invariance at up to $O(10^{-17})$ \cite{SBK2014,SAS2024}. 
In addition, as we show here, 
the nature of rovibrational states of the bound protons makes it possible to isolate the potential 
violation of Lorentz and \textsf{CPT} invariance in the proton (or more generically, hadron) sector
from the combined electron-proton effect observable in atomic spectroscopy. We show here that
this feature alone gives an enhancement of $O(m_p/m_e) \sim 10^3$ in the precision of constraints
on \textsf{CPT} violation from rovibrational transitions with molecular ions compared to those 
possible with atoms alone.

The theoretical framework we use to discuss Lorentz and \textsf{CPT} violation is a low-energy effective 
theory in which the QED Lagrangian is extended to include Lorentz tensor operators with fixed
couplings. These couplings may be thought of as the vacuum expectation values of new tensor
fields which, unlike the familiar case of the scalar Higgs field VEV, necessarily spontaneously break
Lorentz, and in some cases also \textsf{CPT}, symmetry.

This framework is more generally known as the Standard Model Extension (SME) 
\cite{Colladay:1998fq, Kostelecky:2013rta} and for many years
has been used extensively to systematise constraints on Lorentz and \textsf{CPT} symmetry
breaking from a wide variety of experimental data \cite{Kostelecky:2008ts}.  
In the form we use in this paper, the 
Lagrangian for a single Dirac fermion field $\psi(x)$ is extended to include  a set of Lorentz tensor 
operators as follows:
\begin{align}
{\cal L}_{\rm SME} \,=\,& \frac{1}{2}\int d^4 x~\Bigl[ 
\bar\psi\left(i \c^\m \partial_\m \,-\, m\right)\psi
\,-\, a^{}_\m \,\bar\psi \c^\m \psi \,+\, i c^{}_{\m\n}\,\bar\psi \c^\m \partial^\n\psi \,+\,
a^{}_{\m\n\l}\, \bar\psi \c^\m \partial^\n \partial^\l\psi     \nonumber \\[3pt]
& -\, b^{}_\m\, \bar\psi\c^5 \c^\m \psi 
\,+\, i d^{}_{\m\n}\, \bar\psi \c^5\c^\m \partial^\n \psi  
\,-\,  \tfrac{1}{2} H^{}_{\m\n}\, \bar\psi\s^{\m\n}\psi  \,+\, 
\tfrac{1}{2}i g^{}_{\m\n\l}\, \bar\psi\s^{\m\n} \partial^\l \psi
~+~\ldots~ \Bigr] \nonumber \\[3pt]
&+~~ {\rm h.c.}
\label{a1}
\end{align}
To understand the structure of the SME couplings, recall that the standard basis for the 16 possible 
$4\times 4$ matrices acting on the Dirac spinor is $\Gamma = {\bf 1},\, \c^5, \, \c^\m, \,
\c^5 \c^\m, \, \s^{\m\n}$, respectively scalar, pseudoscalar, vector, 
axial vector and tensor. The basic operators then take the form $\bar\psi \Gamma \psi$,
and we can then add increasing numbers of derivatives, which increases the dimension of the operator. 

Restricting to operators with dimension $\le 4$ leaves the theory renormalisable, just like the 
Standard Model itself. This restriction is known as the minimal SME. If on the other hand the
SME is regarded as a low-energy effective theory, valid below some very high energy scale $\Lambda$,
then we may include higher-dimensional operators. The corresponding couplings, such as $a^{}_{\m\n\l}$,
have negative mass dimensions. Here, we keep only the renormalisable couplings in the expansion (\ref{a1})
except for the inclusion of the coupling $a^{}_{\m\n\l}$ of the higher dimensional operator 
$\bar\psi \c^\m \partial^\n \partial^\l\psi$, since this gives the leading spin-independent contribution
to the difference of the $\textrm{H}_2^+$ and $\overline{\textrm{H}}_2^{\,-}$ spectra.
As usual we assume that like the electron, the proton itself is effectively described by this Lagrangian,
with its own distinct couplings, despite it being a bound state of the fundamental quarks, 

In this paper, our main focus is on the `spin-independent' couplings $c_{\m\n}$ and $a_{\m \n \l}$ 
(so-called because in the non-relativistic limit they do not couple to the spin operator; see (\ref{c1})),
though we comment on spin and the hyperfine structure of $\textrm{H}_2^+$ and
$\overline{\textrm{H}}_2^{\,-}$ in section \ref{sect 7}.
These couplings are not observable in spin precession experiments and are therefore much less stringently
constrained than the remaining, spin-dependent, couplings $b_\m, d_{\m\n}, H_{\m\n}$ and $g_{\m\n\l}$.
Indeed the best existing laboratory constraint on the \textsf{CPT} violating coupling $a_{\m\n\l}$ comes 
from the ALPHA antihydrogen $1S$\,-\,$2S$ measurement. 

The SME has important advantages in analysing potential violations of Lorentz and \textsf{CPT} symmetry. 
It provides a systematic parametrisation of possible symmetry breaking effects in terms of a 
standard set of couplings, which allows quantitative comparisons between experiments. 
It also makes it very clear that Lorentz and \textsf{CPT} violation may occur in many different ways
and show up in some experimental measurements while remaining entirely hidden in others.

For example, it is entirely possible for $\textsf{CPT}$ violation to be absent in the antihydrogen 
$1S$\,-\,$2S$ transition, yet appear in transitions such as $1S$\,-\,$2P$ involving states with
non-zero orbital angular momentum, since these involve different SME couplings 
(see section \ref{sect 7}). This is an important motivation to pursue an extensive programme
of high-precision measurements of many different spectral transitions in ${\rm H}$ and
$\overline{\rm H}$, also including searching for sidereal and annual variations of the transition
frequencies which would be a clear indication of Lorentz violation.

On the other hand, the symmetry breaking realised in the SME is comparatively mild and,
as we have mentioned above, can be interpreted as spontaneous symmetry breaking in a theory
which maintains all the essential features of local relativistic QFT, including Lorentz covariance.
In particular, the usual equality of masses of particles and antiparticles (and of course their
identical, but opposite, charges) is maintained in the SME, since the Lagrangian (\ref{a1}) 
is built on the original local causal fields of QED. We should therefore keep an open mind
about more radical alternatives, including non-local theories, even though they generally
lead more immediately to fundamental problems with unitarity and causality 
(see, for example, \cite{Charlton:2020kie}).

The analysis of the molecular ${\rm H}_2^{\,+}$ ion described here is based on the traditional
Bohr-Oppenheimer approximation, in which the 3-body problem is separated into two
Schr\"odinger equations. The first describes the electron wavefunction at fixed nucleon 
separation $R$. The corresponding energy eigenvalues are then interpreted as an
inter-nucleon potential $V_M(R)$ (Fig.~\ref{FigEnergy}) in a second Schr\"odinger equation 
describing the rovibrational motion of the nucleons. The SME is introduced perturbatively
as a non-relativistic Hamiltonian (see (\ref{c1})) derived from the Lagrangian (\ref{a1}).

The application of the SME to the hydrogen molecule and molecular ion has been previously 
studied in \cite{Muller:2004tc} (see also \cite{Kostelecky:2015nma}).  
In this paper, we develop and extend this work in a number of important respects. 
First, we develop a systematic perturbation theory to determine the
rovibrational energy levels and their SME corrections in terms of the potential $V_M(R)$ 
including anharmonic contributions. We see how these are essential to give an accurate
characterisation of the rovibrational energy spectrum from which to evaluate the 
contribution of the SME couplings.

Most importantly, we include the Lorentz and \textsf{CPT} violating couplings for the protons, 
as well as the electron, which were not considered in \cite{Muller:2004tc,Kostelecky:2015nma}.
These have two r\^oles. Along with the electron SME couplings, they modify the inter-nucleon
potential and consequently the rovibrational energies.  The dependence on the electron
and proton SME couplings from this mechanism is identical to that encountered already
in the single atom energy levels. However, the proton couplings also enter directly in
the nucleon Schr\"odinger equation. We show here that the different mass dependence
of the coefficients of the SME couplings in this case results in the $O(m_p/m_e)$
enhancement in sensitivity to \textsf{CPT} violation in the proton sector
highlighted above.

We present our results in terms of the following expansion of the rovibrational energy levels:
\begin{align}
E_{v NM_N} \,=\, \,&V_{\rm SME}^e \,+\, (1 + \d_{\rm SME} )\,(v+\thalf) \,\w_0  
\,\,-\,\, (x_0 + x_{\rm SME} )\,(v+\thalf)^2 \,\w_0 \nonumber \\[5pt]
&+\, (B_0 + B_{\rm SME} ) \,N(N+1) \,\w_0   \, \, -\,\,  (D_0 + D_{\rm SME} ) \,(N(N+1))^2 \,\w_0 
\nonumber \\[5pt]
&\,-\,  (\a_0 +\a_{\rm SME} )\,(v + \thalf) N(N+1) \,\w_0  \,+\, \ldots
\label{a2}
\end{align}
We show that the coefficients here satisfy a hierarchy in terms of the small dimensionless parameter 
$\l = 2/(m_p \,\w_0 \,R_0^2) \,=\, 0.027~$
(where $\w_0$ is the fundamental vibration frequency and $R_0$ is the mean bond length
in the absence of centrifugal and SME corrections),
with $\d_0, \,x_0, \,B_0, \,\a_0, \,D_0$ of order $1, \,\l, \,\l, \,\l^2, \,\l^3$ respectively. 
Each of these coefficients is itself a perturbative expansion in $\l^2$. 

The SME coefficients $\d_{\rm SME},\,B_{\rm SME}, \ldots~$ are themselves the sum of electron and 
proton parts, reflecting the two ways described above in which the SME couplings influence the
rovibrational spectrum. We determine explicitly how these coefficients depend on certain
combinations of the spin-independent SME couplings $c_{\m\n}$ and $a_{\m\n\l}$ in the
Lagrangian (\ref{a1}). In a standard spherical tensor notation \cite{Kostelecky:2013rta}, 
these are denoted
$c_{200}^{{\rm NR}\, w}$, \,$a_{200}^{{\rm NR}\, w}$,  $\,c_{220}^{{\rm NR}\, w}$,  and $\,a_{220}^{{\rm NR}\, w}$, 
where $w = e, p$.  Of these, the $(200)$ couplings are constrained by the two-photon
$1S$\,-\,$2S$ transitions in atomic ${\rm H}$ and $\overline{\rm H}$, while the $(220)$ only 
appear in the single-photon $1S$\,-\,$2P$ transitions for which the natural linewidth is
much broader and the constraint on the SME couplings correspondingly weaker. 
Both may be equally extracted from
rovibrational transition data on the molecular ions ${\rm H}_2^{\,+}$ and $\overline{\rm H}_2^{\,-}$.
The dependence on the $(220)$ couplings introduces the $M_N$ dependence in $E_{v N M_N}$
and shows that even the spin-independent couplings contribute to the hyperfine-Zeeman levels
for $N\neq 0$.

The paper is organised as follows. The Born-Oppenheimer approach is reviewed briefly in 
section \ref{sect 2} and extended to include the SME in section \ref{sect 3}. Section \ref{sect 4}
and Appendix \ref{appendix A} are devoted to solving the electron Schr\"odinger equation in the 
presence of the SME couplings, including numerical evaluations of the relevant expectation values.
In section \ref{sect 5} and Appendix \ref{appendix B} we develop a systematic perturbation theory
for calculating the SME contributions to the rovibrational energies 
for ${\rm H}_2^{\,+}$ in terms of the inter-nucleon potential $V_M(R)$ and its derivatives in two
complementary ways. Section \ref{sect 6} then describes the quantitative effect on the
rovibrational spectrum. 

Finally, in section \ref{sect 7} we take a first look at the rovibrational transitions arising from
$E_{v N M_N}$ in (\ref{a2})  in comparison with analogous results for electron transitions in
single ${\rm H}$ and $\overline{\rm H}$ atoms, and discuss the constraints on the SME
couplings that could be obtained from measurements of rovibrational frequencies with 
${\rm H}_2^{\,+}$  and $\overline{\rm H}_2^{\,-}$, including the search for sidereal and annual 
variations.  We also comment briefly on the hyperfine-Zeeman spectrum, where both the
spin-independent and spin-dependent SME couplings contribute to the energy levels.
This is described in detail in a sequel to this paper \cite{Shore:2025zor}, which we refer to
as Paper 2.

\newpage

\section{Born-Oppenheimer analysis of the hydrogen molecular ion} \label{sect 2}

We begin by reviewing the standard analysis of the spectrum of the hydrogen molecular ion $\textrm{H}_2^+$
in the Born-Oppenheimer approximation, before introducing Lorentz and \textsf{CPT} violation in section
\ref{sect 3}.

The starting point is the 3-body Schr\"odinger equation for the bound state. We denote the positions 
of the nucleons
as $\boldsymbol{r}_1$, $\boldsymbol{r}_2$ and the electron as $\boldsymbol{r}_e$, with corresponding momenta
$\boldsymbol{p}_1$, $\boldsymbol{p}_2$ and $\boldsymbol{p}_e$.  For generality, we temporarily allow 
the masses $m_1$ and $m_2$ to be different in this section. The Hamiltonian is
\begin{equation}
H_{mol} \,=\, \sum_{w=1,2,e} \frac{p_w^2}{2m_w} ~+~ V_{mol}(R, r_{1e}, r_{2e}) \ ,
\label{b1}
\end{equation}
with the electromagnetic potential,
\begin{equation}
V_{mol}(R, r_{1e}, r_{2e}) \,=\,  \a \left(\frac{1}{R} - \frac{1}{r_{1e}} - \frac{1}{r_{2e}}\right) \ ,
\label{b2}
\end{equation}
where $R = |\boldsymbol{r}_1 - \boldsymbol{r}_2|$, $r_{1e} = |\boldsymbol{r}_1 - \boldsymbol{r}_e|$,
$r_{2e}=|\boldsymbol{r}_2 - \boldsymbol{r}_e|$ and $\a$ is the fine structure constant.

Next, we introduce CM variables for the coordinates and momenta. Iterating the standard construction for 
a 2-body system, we define
\begin{align}
\boldsymbol{R} \,&=\,  \boldsymbol{r}_1 - \boldsymbol{r}_2  ~, ~~~~~~~~~~~~~~~~
\boldsymbol{r} \,=\,  \boldsymbol{r}_e  - \frac{1}{M} \, \left(m_1 \boldsymbol{r}_1 
+ m_2 \boldsymbol{r}_2\right)   \nonumber \\
\boldsymbol{R}_{CM} \,&=\, \frac{1}{\hat{M}} \, \left(m_1 \boldsymbol{r}_1 + m_2 \boldsymbol{r}_2 
+ m_3 \boldsymbol{r}_3\right) \ ,
\label{b3}
\end{align}
with $M= m_1 + m_2$ and $\hat{M} = m_1 + m_2 + m_e$, so that $\boldsymbol{R}$ is the inter-nucleon
separation, $\boldsymbol{r}$ is the electron position relative to the nucleon CM, and 
$\boldsymbol{R}_{CM}$ is the position of the CM of the whole molecule.   The corresponding momenta are
\begin{align}
\boldsymbol{P} \,&=\, \mu \dot{\boldsymbol{R}} \,=
\,\frac{1}{M} \, (m_2 \boldsymbol{p}_1 - m_1 \boldsymbol{p}_2 )
~, ~~~~~~~~~~~~~
\boldsymbol{p} \,=\, \hat{\mu}\dot{\boldsymbol{r}} \,=\, \frac{1}{\hat{M}} \big( M \boldsymbol{p}_e 
- m_e (\boldsymbol{p}_1 + \boldsymbol{p}_2 ) \big) \nonumber \\
\boldsymbol{P}_{CM} \,&=\, \hat{M} \dot{\boldsymbol{R}}_{CM} \,=\,
\boldsymbol{p}_1 + \boldsymbol{p}_2 +\boldsymbol{p}_e \ ,
\label{b4}
\end{align}
where we introduce the reduced masses $\mu = m_1 m_2/M$ and $\hat{\mu} = m_e M/\hat{M}$.
The relative motion of the nucleons is then treated as that of a single particle at position $\boldsymbol{R}$ 
with momentum $\boldsymbol{P}$ and reduced mass $\mu$.

With these definitions,\footnote{Notice that with these definitions, in contrast to defining the 
electron momentum relative to the geometric centre of the molecule, there are {\it no} mixed terms 
in the momenta of the form $\boldsymbol{P}.\boldsymbol{p}$ in (\ref{b5}) and 
the Schr\"odinger equation (\ref{b6}) even for a heteronuclear molecule with $m_1 \neq m_2$.}  
the kinetic term in the Hamiltonian (\ref{b1}) becomes simply
\begin{equation}
\sum_{w=1,2,e} \frac{p_w^2}{2m_w} \,=\, \frac{1}{2\mu} P^2 \,+\, \frac{1}{2\hat{\mu}} p^2 \,+\, 
\frac{1}{2\hat{M}} P_{CM}^2 \ .
\label{b5}
\end{equation}
The Schr\"odinger equation for the molecule wavefunction $\Psi(\boldsymbol{R}, \boldsymbol{r},
\boldsymbol{R}_{CM})$ is therefore
\begin{align}
&\left(-\frac{1}{2\mu} \nabla_{\boldsymbol{R}^2} \,-\, \frac{1}{2\hat{\mu}} \nabla_{\boldsymbol{r}^2} \,-\, 
\frac{1}{2\hat{M}} \nabla_{\boldsymbol{R}_{CM}}^2 \,+\, V_{mol}(R, r_{1e}, r_{2e}) \,\right) 
\Psi(\boldsymbol{R}, \boldsymbol{r},\boldsymbol{R}_{CM}) \nonumber \\ 
&~~~~~~~~~~~~~~~~~~~~~~~~~~~~~~~~~~~~~~~~~~~~~~~~~~~~~~~~~~~~~~~~~~~~~~~~~~~=\, 
E \,\Psi(\boldsymbol{R}, \boldsymbol{r},\boldsymbol{R}_{CM}) \ .
\label{b6}
\end{align}
Of course since we are not interested in the bulk motion of the molecule, from now on we set the CM momentum
$\boldsymbol{P}_{CM}$ to zero, and equivalently neglect the $\nabla_{\boldsymbol{R}_{CM}}^2$ term in (\ref{b6}).

The Born-Oppenheimer approximation\footnote{A detailed justification of the Born-Oppenheimer method
may be found in most standard textbooks; see, for example, \cite{QM}. } 
now consists of separating the Schr\"odinger equation into 
two parts.  Writing the molecular wave function $\Psi(\boldsymbol{R},\boldsymbol{r})
= \Phi(\boldsymbol{R})\, \psi(\boldsymbol{r};R)$, we first solve the Schr\"odinger equation
for the electron wavefunction $\psi(\boldsymbol{r};R)$ for fixed nucleon separation $R$.
The energy eigenvalues $E_e(R) \equiv V_M(R)$ then appear as a potential in the effective
Schr\"odinger equation for the nucleons, which determines the rovibrational energy levels of the molecular ion.
We therefore write the ``electron Schr\"odinger equation'',
\begin{equation}
\left(- \frac{1}{2\hat{\mu}} \nabla_{\boldsymbol{r}}^2 \,+\, V_{mol}(R,r_{1e},r_{2e}) \right)\,\psi(\boldsymbol{r};R)
\,=\, E_e(R) \,\psi(\boldsymbol{r};R) \ .
\label{b7}
\end{equation}
We also restrict to the electron ground state $1s\sigma_g$.
Substituting back into (\ref{b6}), the ``nucleon Schr\"odinger equation'' is then
\begin{equation}
\left(-\frac{1}{2\mu} \nabla_{\boldsymbol{R}}^2 \,+\, V_M(R) \right)\, \Phi(\boldsymbol{R}) \,=\, E\, \Phi(\boldsymbol{R})\ .
\label{b8}
\end{equation}
Finally, exploiting the spherical symmetry of the molecular axis in the fixed frame of the experiment
(denoted \textsf{EXP}, see section \ref{sect 3})
to write $\Phi(\boldsymbol{R}) = \frac{1}{R} \phi(R) Y_{N M_N}(\theta,\phi)$ in terms of spherical harmonics,
we find
\begin{equation}
\left( - \frac{1}{2\mu}\, \frac{d^2}{dR^2} \,+\, \frac{1}{2\mu R^2}\, N(N+1) \,+\, V_M(R) \right) \,\phi(R) \,=\,
E_{vN}\, \phi(R) 
\label{b9}
\end{equation}
Here, we follow the widely-used convention of denoting the discrete vibrational energy states by the
integer $v$ and the nucleon angular momentum states by $N, M_N$, with $M_N$ the $3$-component with
respect to the \textsf{EXP} frame. 

An approximate solution to the electron Schr\"odinger equation, based on an $R$-dependent {\it ansatz} for the 
wavefunction $\psi(\boldsymbol{r};R)$, is discussed in detail in Appendix \ref{appendix A}.
The method is essentially standard, but we require some special features and numerical results 
which ultimately feed into the coefficients of the Lorentz and \textsf{CPT} violating couplings 
constrained by the rovibrational spectrum of the (anti-)molecular ion.  Inserting the resulting 
$R$-dependent energy eigenvalues into the nucleon Schr\"odinger equation then gives an inter-nucleon 
potential of the typical Morse potential form illustrated in Fig.~\ref{FigEnergy} in section \ref{sect 4}.

For $\textrm{H}_2^+$, this has a minimum at $R_0 \simeq 2 a_0$  (where $a_0$ is the Bohr radius)
about which the nucleons undergo approximately simple harmonic motion with angular frequency $\w_0$
and integer vibrational quantum number $v$.  
For $\textrm{H}_2^+$, $\w_0 \simeq 0.02\,R_H$, where $R_H$ is the Rydberg energy, $13.6$\,eV. 
The spherical symmetry implies that the rovibrational energy levels depend only
on the nucleon orbital angular momentum quantum number $N$, and {\it not} the component $M_N$.
As we see in the next section, however, this is no longer true when the Lorentz and \textsf{CPT} 
violating interactions are introduced.

Solving the nucleon Schr\"odinger equation then allows the
rovibrational energy levels $E_{v N}$ to be
written as an expansion in $(v + \tfrac{1}{2})$ and $N(N+1)$, {\it viz.}\footnote{The notation here is relatively
standard (see {\it e.g.} \cite{Varshalovich}), but note that we have taken out a common energy factor $\w_0$
so the coefficients $B_0, x_0, \a_0, D_0, \ldots$ here are all dimensionless.}
\begin{align}
E_{v N} \,=\, &(v+\thalf) \w_0  \,-\,  x_0 (v+\thalf)^2 \w_0 \nonumber \\
&+\, B_0 \,N(N+1) \,\w_0   \,-\,  \a_0 \,(v + \thalf) N(N+1)\,\w_0  \,-\,  D_0\, (N(N+1))^2\,\w_0  \,+\,  \ldots \ ,
\label{b10}
\end{align}
where $\w_0^2 = V_M^{''}(R_0)/\mu$. 
It will be very useful in organising this expansion to introduce the small dimensionless parameter 
$\l = 1/(\mu\, \w_0 R_0^2)$, which for $\textrm{H}_2^+$ is $ \l \simeq 0.027$. 
 From (\ref{b9}) it follows directly that at leading order, $B_0 = \l/2$.  
In fact, all the coefficients are themselves power series in $\l$, determined by higher derivatives of the potential $V_M(R)$,
with their leading terms displaying a hierarchy in this small parameter. In section \ref{sect 5}, we show explicitly
that the leading terms for $x_0$, $\a_0$ and $D_0$ are of order $\l, \l^2$ and $\l^3$ respectively.

Explicit analytic expressions for these coefficients will be given later, and evaluated numerically to a precision 
sufficient to determine the prefactors of the contributions of the Lorentz and \textsf{CPT} violating couplings to the 
rovibrational energy levels.  Of course, extensive calculations in high-order QED carried out over many years 
have determined them to extremely high precisions of a few parts in  $10^{12}$, enabling direct comparisons
of experiment with {\it ab intio} theory (see {\it e.g.}~\cite{Korobov:2017}), 
but this sort of precision is not needed for our purpose here.

\section{Born-Oppenheimer analysis with Lorentz and \textsf{CPT} violation}\label{sect 3}

In this section, we extend the Born-Oppenheimer analysis of the $\textrm{H}_2^+$ molecular ion to include the effects 
of possible Lorentz and \textsf{CPT} violation.  

We start from the non-relativistic Hamiltonian derived from the SME Lagrangian (\ref{a1}) for a single
Dirac fermion:
\begin{equation}
H^{\rm NR}_{\rm SME} ~=~ \bigl(A\,+2\,B_k S^k\bigr) \,+\, \bigl(C_i \,+\, 2D_{ik} S^k\bigr)\frac{p^i}{m} 
\,+\, \bigl(E_{ij} \,+\, 2F_{ijk} S^k\bigr) \frac{p^i p^j}{m^2} \ ,
\label{c1}
\end{equation}
where $S^k$ is the spin operator.  Since QED conserves parity, the leading perturbative contributions 
to expectation values from
$C_i$ and $D_{ij}$ vanish, while $A$ gives only a common addition to the energy levels and does not affect
the spectrum. In terms of the fundamental SME couplings, the relevant coefficients for spectroscopy are
\cite{Kostelecky:1999zh, Yoder:2012ks, Kostelecky:2013rta}
\begin{align}
B_k ~&=~ - b_k \,+\, m\, d_{k0} \,+\, \tfrac{1}{2}\e_{kmn}
\big(H_{mn} - m\, g_{mn0}\big) \ ,
\nonumber \\[5pt]
E_{ij} ~&=~ -m\big(c_{ij} \,+\, \tfrac{1}{2}\, c_{00}\, \d_{ij}\big) \,+\,
m^2 (3\, a_{0ij} \,+\, a_{000}\, \d_{ij} )  \ ,
\nonumber \\[5pt]
F_{ijk} ~&=~ \tfrac{1}{2}\big(b_k \,\d_{ij} \,-\, b_j \,\d_{ik}\big) \,+\, 
m\,\big(d_{0j} \,+\, \tfrac{1}{2}\,d_{j0}\big) \d_{ik} \,-\, 
\tfrac{1}{4} \d_{ik} \e_{jmn}\, H_{mn} \nonumber \\[2pt]
&~~~~~~~~~~~~~~~~~~~~~~~~~~~~~~~~~~~~~~~~~~~~~~~~~~~~~~~~
\,-\, m\, \e_{ikm} \big(g_{m0j} + \tfrac{1}{2}\, g_{mj0}\big)   \ .
\label{c2}
\end{align}

To keep the presentation reasonably simple, and because the spin-dependent couplings are experimentally
already far more constrained than the spin-independent couplings, we discuss mainly the spin-independent 
effects in this paper. 
These arise from the coupling combinations $E^p_{ij}$, $E^e_{ij}$ for the protons and electron respectively.
The analysis of the spin-dependent contributions is then conceptually largely straightforward and will be 
briefly introduced in section \ref{sect 7}, where we discuss some aspects of the spectrum including 
hyperfine-Zeeman splittings.

With the inclusion of the Lorentz violating couplings, we need to be especially careful in specifying the 
reference frame in which the components are defined. Altogether, there are four frames of reference 
which are relevant to our discussion of the molecular ion.
In order to compare constraints on the SME couplings between different experiments it has become standard 
practice to quote bounds ultimately in terms of a standard Sun-centred frame (\textsf{SUN}), which is
in turn related to a ``standard laboratory frame'' (\textsf{LAB}). The precise definitions of these standard frames 
and the coordinate transformations relating them may be found, for example, in \cite{Kostelecky:2015nma}.

Here, we are mainly concerned with two further frames.  An ``apparatus frame'' (\textsf{EXP}), with basis
vectors $\boldsymbol{e}_i$\, ($i$ = 1,2,3), is chosen which is specific to the particular experiment and, 
importantly, in which the angular momentum components considered below are defined. Typically it will be
chosen such that the $\boldsymbol{e}_3$ basis vector is aligned with a background magnetic field.
We define the nucleon rotational quantum numbers $N,M_N$ with respect to this \textsf{EXP} frame.
Importantly, the SME couplings $E_{ij}$ are, as indicated by the indices, also expressed here 
in the \textsf{EXP} frame.

Finally, in particular in the solution of the electron Schr\"odinger equation, we will also work in a frame 
(\textsf{MOL}) with basis vectors $\boldsymbol{e}_a$\, ($a =x,y,z$) with the $z-$axis 
aligned with the inter-nucleon, or molecular, axis.

To implement the Born-Oppenheimer analysis in this theory, we first rewrite the SME Hamiltonian 
in terms of the CM momenta introduced above.  
For $\textrm{H}_2^+$, the reduced masses are $\mu = m_p/2$ and $\hat{\mu} 
= m_e \left(1 + m_e/2m_p\right)^{-1}$
and we find, keeping only the $E_{ij}$ couplings,
\begin{equation}
\D H_{\textrm{SME}} \,=\, \frac{2}{m_p^2} E_{ij}^p \, P^i P^j  \,+\, \Bigl( \frac{1}{2 m_p^2} E_{ij}^p 
\,+\, \frac{1}{m_e^2} E_{ij}^e\Bigr) p^i p^j \ .
\label{c3}
\end{equation}
Notice that there is no mixed momentum term proportinal to $P^i p^j$ in this SME 
Hamiltonian. This is a special feature of the homonuclear case, where $m_1=m_2$, and is 
not true for heteronuclear molecular ions such as $\textrm{HD}^+$.  

Splitting the Schr\"odinger equation for the molecule in the Born-Oppenheimer approximation as above,
and evaluating throughout in the \textsf{MOL} frame, we therefore find
\begin{equation}
\left(- \frac{1}{2\hat{\mu}} \nabla_{\boldsymbol{r}}^2 \,+\, V_{mol}(R,r_{1e},r_{2e}) 
+ \Bigl( \frac{1}{2 m_p^2} E_{ab}^p \,+\, \frac{1}{m_e^2} E_{ab}^e\Bigr)p^a p^b \right)\,\psi(\boldsymbol{r};R)
\,=\, E_e(\boldsymbol{R}) \,\psi(\boldsymbol{r};R) \ ,
\label{c4}
\end{equation}
where we understand $\boldsymbol{p} \rightarrow -i \nabla_{\boldsymbol{r}}$.
We write $E_e(\boldsymbol{R}) \equiv  V_M(R) \,+\, V_{\textrm{SME}}^e(\boldsymbol{R})$ here to remember that
$V_{\textrm{SME}}^e(\boldsymbol{R})$ depends on the orientation $(\theta,\phi)$ of the molecular axis in the
\textsf{EXP} frame, but through the SME couplings alone.  The relation of these couplings 
in the \textsf{MOL} and \textsf{EXP} frames is calculated in the next section.

The nucleon Schr\"odinger equation, which is expressed in the \textsf{EXP} frame, is then
\begin{equation}
\left(-\frac{1}{2\mu} \nabla_{\boldsymbol{R}}^2 \,+\, V_M(R) \,+\, V_{\textrm{SME}}^e(\boldsymbol{R}) 
\,+\, \frac{2}{m_p^2} E_{ij}^p P^i P^j \right)\, \Phi(\boldsymbol{R}) \,=\, E_{vNM_N}\, \Phi(\boldsymbol{R}) \ .
\label{c5}
\end{equation}
We see here that with the Lorentz and \textsf{CPT} breaking term $V_{\textrm{SME}}^e(\boldsymbol{R})$,
the usual spherical symmetry of the unperturbed nucleon Schr\"odinger equation is broken.
In turn, the degeneracy of the rovibrational energy levels at fixed $N$ is broken and they acquire a 
dependence also on the quantum number $M_N$.  

Notice also how, perhaps unexpectedly, the proton SME couplings appear already in the Schr\"odinger 
equation for the electron and so modify the effective potential for the nucleon motion as well as their 
direct appearance in (\ref{c5}).  To simplify notation, from now on we use the abbreviated notation
notation $\tilde{E}_{ij}^e = E_{ij}^e + \thalf (m_e^2 / m_p^2)\, E_{ij}^p$. Note that despite the small pre-factor,
we should not immediately drop the second term as we have no {\it a priori} knowledge of the
relative sizes of the SME couplings for different particles (see also section \ref{sect 7}).

Next, expressing $\Phi(\boldsymbol{R})$ in terms of spherical harmonics as before, we may write 
the analogue of (\ref{b9}) as follows:
\begin{equation}
\left( - \frac{1}{2\mu}\, \frac{d^2}{dR^2} \,+\, \frac{1}{2\mu R^2} N(N+1) \,+\, V_M(R) \,+\,
V_{\textrm{SME}}^e(R) \right) \phi(R)  
=\, \tilde{E}_{vNM_N}\, \phi(R) \ ,
\label{c6}
\end{equation}
where
\begin{equation}
V_{\textrm{SME}}^e(R) \,=\, \langle N M_N |\, V_{\textrm{SME}}^e(\boldsymbol{R})\,|N M_N\rangle \,=\,
\int d\Omega \,Y_{NM_N}^*(\theta,\phi) \,\, V_{\textrm{SME}}^e(\boldsymbol{R}) \,\,Y_{NM_N}(\theta,\phi)  \ .
\label{c7}
\end{equation}
The contribution of the $E_{ij}^p$ term in (\ref{c5}) to the total rovibrational energies $E_{v N M_N}$ 
is calculated in first order perturbation theory by evaluating $\langle P^i P^j\rangle$ in the unperturbed
nucleon state.  Defining
 \begin{equation}
\Delta E_{\textrm{SME}}^n \,=\, \frac{2}{m_p^2} E_{ij}^p \,\langle v\, N M_N |\, P^i P^j \,|v\, N M_N\rangle \ ,
\label{c8}
\end{equation}
we finally have
\begin{equation}
E_{v N M_N} \,=\,  \tilde{E}_{v N M_N} \,+\,  \Delta E_{\textrm{SME}}^n 
\label{c9}
\end{equation}
In the following sections, we solve these equations and evaluate the SME corrections
to the rovibrational energy levels.

\section{Electron Schr\"odinger equation and inter-nucleon potential}\label{sect 4}

We are now in a position to determine the inter-nucleon potential $V_M(R)$ by solving the electron Schr\"odinger
equation (\ref{b7}).  We then include the Lorentz and \textsf{CPT} breaking couplings and evaluate the
potential $V_{\textrm{SME}}^e(R)$ in (\ref{c6}) and its consequent effect on the rovibrational energy levels.

The detailed calculations of the energy eigenvalues and momentum expectation values in the absence
of Lorentz and \textsf{CPT} breaking are described in Appendix \ref{appendix A}, so here we just summarise 
the essential  features. We start from the following {\it ansatz} for the electron wave function,
as always in the $1s\sigma_g$ ground state,
\begin{equation}
\psi(\boldsymbol{r};R) \,=\, \frac{1}{\sqrt{2(1 + I_0(R))}} \left( \psi_H(r_{1e};R) \,+\, \psi_H(r_{2e};R)\right) \ ,
\label{d1}
\end{equation}
where 
\begin{equation}
\psi_H(r;R) \,=\, \sqrt{\frac{\c(R)^3}{\pi \hat{a}_0^3}}\, e^{- \c(R) r/\hat{a}_0} \ .
\label{d2}
\end{equation}
and the overlap function $I_0(R)$ ensures the wavefunction is correctly normalised. 
The $\psi_H(r;R)$ are hydrogen $1s$ wavefunctions with reduced Bohr radius $\hat{a}_0$, modified by
an interpolating function $\c(R)$ which adjusts the effective Bohr radius according to the inter-nucleon
separation $R$ \cite{Muller:2004tc}. 
This function is determined numerically by minimising the energy eigenvalue $E_e(R)$
for each value of $R$. It is also constrained physically to take the values $\c(0) = 2$ and 
$\c(R)\rta 1$ for large $R$. This ensures the wavefunction reduces to that of a hydrogen-like atom
with $Z=2$ as the nucleon separation goes to zero, while for large separations the molecule effectively
separates into an isolated proton and a single hydrogen atom in the $1s$ state with the usual Bohr radius.

With $\c(R)$ set to 1, the energy eigenvalues may be calculated analytically in terms of elementary integrals,
and we find a simple expression for $E_{e1}(R) = K_1(R)  + U_1(R)$ in terms of the corresponding 
kinetic and potential energies, with
\begin{equation}
K_1(R) \,=\,   \frac{1}{(1 + I_0)} \,\Bigl(1 + (1+R-\tfrac{1}{3} R^2) e^{-R}\Bigr) \ ,    
\label{d3}
\end{equation}
and
\begin{equation}
U_1(R) \,=\,   \frac{2}{R}  \,-\, \frac{2}{(1 + I_0)}\, \Bigl( 1 + \frac{1}{R} + 2(1+R)e^{-R} - 
\bigl(1 + \frac{1}{R}\bigr)e^{-2R} \Bigr) \ ,
\label{d4}
\end{equation}
where $I_0(R) = \bigl(1 + R + \tfrac{1}{3}R^2\bigr) \exp(-R)$.\footnote{Here we are using atomic units
(see Appendix \ref{appendix A}) where $R$ is rescaled by the reduced Bohr radius $\hat{a}_0$ and is dimensionless
and energies are similarly scaled by the reduced Rydberg constant$\hat{R}_H$. }
Reinstating  $\c(R)$, the energy $E_e(R)$ may be deduced from these results by inspection, following through
the appropriate rescalings. This gives
\begin{equation}
E_e(R) \,=\,  \c(R)^2 \,K_1(\c(R) R) \,+\,  \c(R)\, U_1(\c(R)R) \ .
\label{d5}
\end{equation}
Inserting the interpolating function $\c(R)$ found numerically in Appendix \ref{appendix A},
$E_e(R)$ is plotted in Fig.~\ref{FigEnergy}. 
As anticipated, it takes the characteristic Morse-like form,
with a minimum at $R_0 = 2.003\, \hat{a}_0$. For small $R$, it is dominated by the inter-nucleon
repulsion $2/R$ while otherwise $E_e(0) \rta -4 \hat{R}_H$, as appropriate for an atom with $Z=2$.
For large $R$ we recover $E_e(R) \rta E_{1s} = - \hat{R}_H$, the ground state energy of a hydrogen atom.
\begin{figure}[h!]
\centering{\includegraphics[scale=0.67]{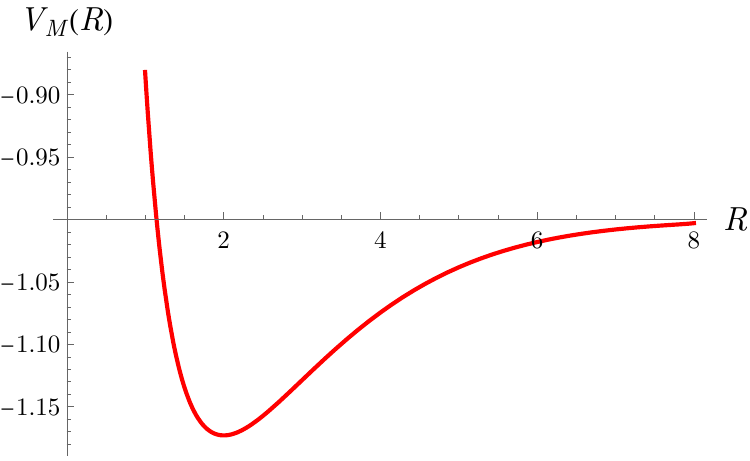}} 
\caption{The nucleon potential $V_M(R) \equiv E_e(R)$ as the inter-nucleon distance, or bond length,
 $R$ is varied. The minimum is at $R_0 = 2.003$ in units of the reduced Bohr radius $\hat{a}_0$. }
\label{FigEnergy}
\end{figure}

As explained above, in the Born-Oppenheimer approximation, the energy eigenvalue $E_e(R)$ is identified
as the inter-nucleon potential $V_M(R)$.  Its curvature at the minimum gives the fundamental vibrational
frequency $\w_0$ of the molecule.  We find
\begin{equation}
\omega_0 =\sqrt{\frac{1}{\mu}  V_M^{''}(R_0)  } = 0.020\,\hat{R}_H \ ,
\label{d6}
\end{equation}
that is, $\w_0=0.275$\,eV.

The expectation values for the momentum, required to evaluate the potential $V_{\textrm{SME}}^e(R)$
are given by 
\begin{equation}
\langle \,p^a \,p^b\,\rangle \,=\, -\int d^3\boldsymbol{r}\, \psi(\boldsymbol{r};R) \,\nabla^a \nabla^b \,
\psi(\boldsymbol{r};R) \ ,
\label{d7}
\end{equation}
where we are working in the \textsf{MOL} frame where the $z$-axis is aligned with the molecular axis.
Cylindrical symmetry then implies $\langle \,p^x p^x\,\rangle = \langle \,p^y p^y\,\rangle$, while 
$\langle p^a p^b\rangle$ vanishes for $a\neq b$. WIth $\c(R)$ set to 1, we find the analytic expressions
\begin{equation}
\langle\,p^x p^x\,\rangle_1\,=\, \langle p^y p^y \rangle_1 \,=\, 
\frac{1}{3} \frac{1}{(1 + I_0)}\, \bigl( 1 + (1+R) e^{-R}\bigr) \ ,
\label{d8}
\end{equation}
and
\begin{equation}
\langle\,p^z p^z\,\rangle_1 \,=\, \frac{1}{3} \frac{1}{(1 + I_0)}\, \bigl( 1 + (1+R-R^2) e^{-R}\bigr) \ .
\label{d9}
\end{equation}
Reinstating $\c(R)$, in this case we have
\begin{equation}
\langle\, p^a p^b\,\rangle_\c(R) \,=\, \c(R)^2 \,\langle\,p^a p^b\,\rangle_1(\c(R) R) \ .
\label{d10}
\end{equation}
These momentum expectation values are plotted in Fig.~\ref{FigmomentaText} 
(see also Fig.~\ref{Figmomenta} in Appendix \ref{appendix A}.  
Since full spherical symmetry is restored as both
$R\rta 0$ and $R\rta \infty$, we expect all three expectation values to be equal in these limits
and related in the obvious way to the kinetic energy, implying (no sum
on $a$)
$\langle p^a \,p^a\rangle \rta 4/3$  as $R\rta 0$ and 
$\langle p^a \,p^a\rangle \rta 1/3$ for large $R$.
This is confirmed in the explicit numerical solutions.
\begin{figure}[h!]
\centering{{\includegraphics[scale=0.59]{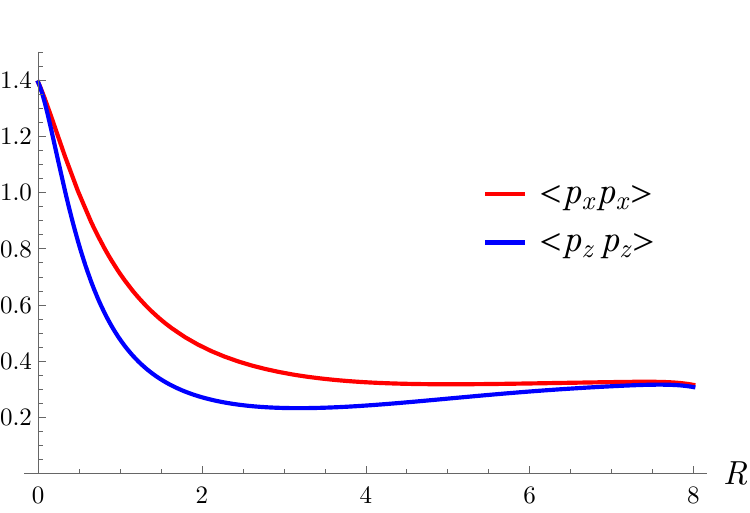}} \hskip0.5cm
{\includegraphics[scale=0.55]{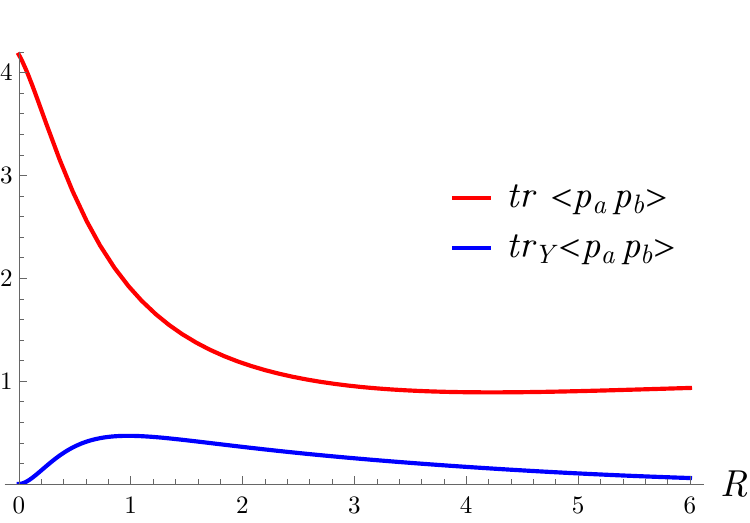}} }
\caption{The momentum expectation values which determine 
the Lorentz and \textsf{CPT} violating contributions to the nucleon Schr\"odinger equation arising
from the SME couplings of the electron as a function of the bond length $R$. 
 The left hand figure shows $\langle p^x\,p^x\rangle$ (red,upper) and  $\langle p^z\,p^z\rangle$ 
(blue) evaluated with the scaling factor $\gamma(R)$.  The right hand figure is the same 
for $\tr\,\langle p^a\,p^b\rangle$ (red) and $\tr_Y\langle p^a\,p^b\rangle$ (blue) defined below.}
\label{FigmomentaText}
\end{figure}

Next, since these momentum expectation values are evaluated in the \textsf{MOL} frame whereas we
want to express the rovibrational energies in terms of the Lorentz and \textsf{CPT} couplings
$\tilde{E}_{ij}$ in the \textsf{EXP} frame, we need to find the relation between these frames.

The rotation matrix relating the \textsf{EXP} and \textsf{MOL} basis vectors, where 
$\boldsymbol{e}_i = \boldsymbol{e}_a\, R_{ai}$, is\footnote{To transform from the \textsf{MOL} 
to the \textsf{EXP} frame we use the usual Euler angles, noting that only two of the three required
in general are necessary here because of the cylindrical symmetry in the \textsf{MOL} frame
of reference.
Then $R_{ai}$ comprises (i) a rotation of $-\theta$ about the \textsf{MOL} $y$-axis, followed by
(ii) a rotation of $-\phi$ about the new \textsf{MOL} $z$-axis, that is
\begin{equation*}
R_{ai} \,=\, \begin{pmatrix}
\cos\theta &~~~ 0 &~~~ -\sin\theta \,\, \\
0 &~~~ 1 &~~~ 0 \\
\sin\theta &~~~ 0 &~~~ \cos\theta 
\end{pmatrix}\, \begin{pmatrix}

\cos\phi &~~~ \sin\phi &~~~ 0 \,\,  \\
-\sin\phi &~~~ \cos\phi &~~~ 0  \\
0 &~~~ 0 &~~~ 1 
\end{pmatrix}  \ .
\end{equation*}  }
\begin{equation}
R_{ai} \,=\, \begin{pmatrix}
\cos\theta \,\cos\phi &~~~ \cos\theta\,\sin\phi &~~~-\sin\theta  \,\,\\
-\sin\phi &~~~ \cos\phi &~~~0 \\
\sin\theta\,\cos\phi &~~~\sin\theta\,\sin\phi &~~~\cos\theta 
\end{pmatrix} \ .
\label{d11}
\end{equation}
The angles $(\theta,\phi)$ here are the standard spherical polar coordinates specifying the 
orientation of the molecular axis in the $\textsf{EXP}$ frame.

The SME couplings are then related by
\begin{equation}
\tilde{E}^e_{ab} \,=\, R_{ai}\, \tilde{E}^e_{ij} \, (R^{\textsf{T}})_{jb}  \ .
\label{d12}
\end{equation}

The SME potential $V_{\textrm{SME}}^e(\boldsymbol{R}) \,\equiv\, V_{\textrm{SME}}^e(R,\theta,\phi)$ in 
(\ref{c5}) is therefore 
\begin{align}
V_{\textrm{SME}}^e(R,\theta,\phi) \,&=\, \frac{1}{m_e^2}\, \langle \,p^a\,p^b\,\rangle\, \tilde{E}^e_{ab} \nonumber \\
&=\, \frac{1}{m_e^2}\, \langle \,p^a\,p^b\,\rangle\, R_{ai}\, \tilde{E}^e_{ij} \, (R^{\textsf{T}})_{jb} 
\label{d13}
\end{align}
and using the cylindrical symmetry to set $\langle p^yp^y\rangle = \langle p^x p^x\rangle$, we can rewrite
this as 
\begin{equation}
V_{\textrm{SME}}^e(R,\theta,\phi) \,=\, \frac{1}{m_e^2}\,  \Big[\,\langle\,p^x\,p^x\,\rangle \,\d_{ij} 
~+~ \bigl(\langle\,p^z\,p^z\,\rangle \,-\, \langle\,p^x\,p^x\,\rangle \bigr) \,
(R^{\textsf{T}})_{iz}\, R_{zj} \, \Big]\, \tilde{E}^e_{ij} \ .
\label{d14}
\end{equation}
This greatly simplifies the calculation, since we now only need to evaluate the symmetric matrix
\begin{equation}
(R^{\textsf{T}})_{iz}\, R_{zj} \,=\, \begin{pmatrix}
\sin^2\theta\,\cos^2\phi &~~~ \tfrac{1}{2}\sin^2\theta\,\sin2\phi &~~~ \tfrac{1}{2} \sin2\theta\,\cos\phi \,\, \\
\ldots &~~~ \sin^2\theta\,\sin^2\phi &~~~ \tfrac{1}{2} \sin2\theta\,\sin\phi  \\
\ldots &~~~ \ldots &~~~ \cos^2\theta 
\end{pmatrix} \ .
\label{d15}
\end{equation}

The final step to derive the potential $V_{\textrm{SME}}^e(R)$ to be used in the nucleon Schr\"odinger equation
(\ref{c6}) is to take the expectation value in (\ref{c7}), {\it viz.}
\begin{equation}
V_{\textrm{SME}}^e(R) \,=\, \int d\Omega \,Y_{NM_N}^*(\theta,\phi) \, V_{\textrm{SME}}^e(R,\theta,\phi) \,
Y_{NM_N}(\theta,\phi)  \ .
\label{d16}
\end{equation}
Notice that this assumes the states are eigenstates of both $N$ {\it and} $M_N$. This will not necessarily
be the case when we consider the hyperfine-Zeeman spectrum, 
which we discuss in detail in Paper 2 \cite{Shore:2025zor}.

To evaluate (\ref{d16}), first expand $(R^{\textsf{T}})_{iz}\, R_{zj}$ in terms of spherical harmonics as
\begin{equation}
(R^{\textsf{T}})_{iz}\, R_{zj} \,=\, \sum_M\, C_{ij}^M\, Y_{2M}(\theta,\phi) \,+\, \tilde{C}_{ij}^0 \,Y_{00}  \ ,
\label{d17}
\end{equation}
with 
\begin{equation}
C_{ij}^M \,=\, \int d\Omega \, (R^{\textsf{T}})_{iz}\, R_{zj}\, Y_{2M}(\theta,\phi) \ .
\label{d18}
\end{equation}
The combination $\tilde{C}_{ij}^0 \,Y_{00} = 1/3 \,\d_{ij}$ follows without further calculation by noting
that $\tr\, (R^{\textsf{T}})_{iz}\, R_{zj} = 1$ and $\tr\, C_{ij}^M = 0$.
In fact, we only need to calculate $C_{ij}^M$ for $M=0$, and with standard normalisations of the spherical
harmonics we readily find \cite{Yoder:2012ks},
\begin{equation}
C_{ij}^0 \,=\, -\frac{2}{3} \sqrt{\frac{\pi}{5}} \, \begin{pmatrix}
1~~~&0~~~&0 \,\, \\
0~~~&1~~~&0 \\  
0~~~&0~~~&-2 
\end{pmatrix}  \ .
\label{d19}
\end{equation}

The expectation value (\ref{d16}) is evaluated in terms of Clebsch-Gordan coefficients using the Gaunt
integral,
\begin{equation}
\int d\Omega \, Y_{NM_N}^*(\theta,\phi) \, Y_{2M}(\theta,\phi) \, Y_{NM_N}(\theta,\phi)  \,=\,
\frac{1}{2} \sqrt{\frac{5}{\pi}} \, \,C_{NM_N;20}^{NM_N} \,\, C_{N0;20}^{N0} \,\, \d_{M0} \ ,
\label{d20}
\end{equation}
and we find
\begin{align}
V_{\textrm{SME}}^e(R) \,=\, \,&\frac{1}{m_e^2}\,  \Big[\,\langle\,p^x\,p^x\,\rangle \,\d_{ij} \nonumber \\
~&+~ \bigl(\langle\,p^z\,p^z\,\rangle \,-\, \langle\,p^x\,p^x\,\rangle \bigr) \,
 \, \Big(\frac{1}{2} \sqrt{\frac{5}{\pi}} \, C_{NM_N;20}^{NM_N} \, C_{N0;20}^{N0} \, C_{ij}^0\,\,
+ \frac{1}{3}\,\d_{ij} \Big) \Big]\, \tilde{E}^e_{ij} \ .
\label{d21}
\end{align}
The Clebsch-Gordan coefficients are known analytically, and simplifying we have
\begin{equation}
C_{NM_N;20}^{NM_N} \, C_{N0;20}^{N0} \,=\, \frac{N(N+1) - 3 M_N^2}{(2N-1)(2N+3) } ~\equiv~ c_{N M_N} \ ,
\label{d22}
\end{equation}
which vanishes for $N=0$. 
Collecting these results and rearranging, we therefore find our final result for the Lorentz and \textsf{CPT}
violating contribution $V_{\textrm{SME}}^e(R)$ to the inter-nucleon potential \cite{Muller:2004tc},
\begin{align}
V_{\textrm{SME}}^e(R) \,=\, \,\frac{1}{3}\frac{1}{m_e^2}\,  \Big[
&\,\big(2 \langle\,p^x\,p^x\,\rangle \,+\langle\,p^z\,p^z\,\rangle \big)\,\, \tr\,\tilde{E}^e_{ij}  \nonumber \\
~&+~ \bigl(\langle\,p^x\,p^x\,\rangle \,-\, \langle\,p^z\,p^z\,\rangle \bigr) \,
 \, c_{N M_N}\,\, \tr_Y \,\tilde{E}^e_{ij}\Big]   \ ,
\label{d23}
\end{align}
where we define $\tr_Y \,\tilde{E}_{ij} \,=\, \tilde{E}_{11}  + \tilde{E}_{22} - 2\tilde{E}_{33}$. 
As anticipated, the Lorentz and \textsf{CPT} violating couplings introduce an explicit $M_N$ dependence
in $V_{\textrm{SME}}^e(R)$\footnote{Logically, we should display the $N,M_N$ dependence
explicitly by writing $V_{\textrm{SME};NM_N}^e(R)$ as we do for the energy eigenvalues $E_{vNM_N}$
but this notation becomes cumbersome and we suppress these labels in what follows.} ultimately breaking 
the degeneracy of the nucleon rovibrational levels with fixed $v,N$.

\section{Rovibrational energy levels}\label{sect 5}

We now move on to an explicit derivation of the rovibrational energy levels for the (anti-)hydrogen
molecular ion, first without Lorentz and \textsf{CPT} violation and then incorporating the breaking
potential $V_{\textrm{SME}}^e(R)$. We work analytically throughout, quoting our final answers in terms
of derivatives of $V_M(R)$ and $V_{\textrm{SME}}^e(R)$ with respect to the inter-nucleon separation $R$,
the latter being governed by the derivatives of the momentum expectation values in (\ref{d21}).
Numerical estimates for these derivatives have been calculated in Appendix \ref{appendix A}, but
we delay using these to keep maximum generality for as long as possible. 
An alternative approach, potentially more systematic, to calculating the rovibrational levels is 
given in Appendix \ref{appendix B}.  Both approaches bring different physical insights into the 
physical origin of the various contributions to the energy levels.

We now pick up the discussion from the end of section \ref{sect 2}. The first objective is to calculate
the coefficients $x_0, B_0, \alpha_0$ and $D_0$ in the theory with no Lorentz or \textsf{CPT} violation.
We start by including the angular momentum term in the nucleon Schr\"odinger equation (\ref{b9})
in an {\it effective} potential $V_{\it eff}(R)$ and study the simple harmonic motion about the minimum
of this potential. So we define
\begin{equation}
V_{\it eff}(R) \,=\,  V_M(R) \,+\, \frac{1}{2 \mu R^2}\,N(N+1)  \ .
\label{e1}
\end{equation}
The angular momentum term changes the mean inter-nucleon separation (bond length) and the 
vibration frequency perturbatively in the small parameter $\l = 1/(\mu \,\w_0 R_0^2)$ that was introduced 
in section \ref{sect 2}.

First, let $R_m = R_0 + \d R$ be the minimum of the effective potential. Setting $V_{\it eff}^{'}(R_m) = 0$,
we find 
\begin{equation}
\frac{\d R}{R_0} \,=\, \l^2 \,N(N+1) \,-\, \Big(3 \,+\, \frac{1}{2} \frac{R_0 V_M^{'''}(R_0)}{V_M^{''}(R_0)}\Big) \,\l^4\,
(N(N+1))^2   \, +\,  \ldots   
\label{e2}
\end{equation}
The corresponding shift in the bond length is interpreted as ``centrifugal stretching'' as the 
centrifugal force due to the orbital angular momentum stretches the effective spring binding the nucleons. 

Evaluating the effective potential at the new minimum gives
\begin{align}
V_{\it eff}(R_m) \,&=\, V_{\it eff}(R_0) \,+\, \d R\, V_{\it eff}'(R_0) \,+\, \frac{1}{2} \d R^2 \, V_{\it eff}^{''}(R_0) \,+\, \ldots
\nonumber \\
&= V_M(R_0) \,+\, \frac{1}{2} \l \,N(N+1)\,\w_0 \,-\, \frac{1}{2} \l^3 \,(N(N+1))^2 \,\w_0 \,+\, \ldots 
\label{e3}
\end{align}
The coefficient of this $O(N(N+1))^2$ term is of importance later. It arises here as a combination of two terms,
one with a pre-factor $-1$ from the $O(\d R)$ contribution, where it is interpreted as reduced
\textit{kinetic} energy due to centrifugal stretching at fixed $N$, and one with pre-factor 1/2 from the 
$O(\d R)^2$ contribution which is interpreted as the extra \textit{potential} energy due to this stretching of the 
inter-nucleon bond.

Next, we need the change in the effective vibrational frequency due to expanding about the minimum
of $V_{\it eff}(R)$. Here,
\begin{align}
V_{\it eff}^{''}(R_m) \,&=\, V_{\it eff}^{''}(R_0)  \,+\, \d R\,V_{\it eff}^{'''}(R_0) \,+\, \ldots \nonumber \\
&=\, V_M^{''}(R_0)\,\Big[\, 1 \,+\, \Big(3 \,+\, \frac{R_0 V_M^{'''}(R_0)}{V_M^{''}(R_0)}\,\Big) \,\l^2 \,N(N+1) 
\,+\, O(\l^4\ (N(N+1))^2) \, \Big] \ .
\label{e4}
\end{align}
We have also calculated the $O(\l^4\, (N(N+1))^2)$ term here, which has a coefficient involving up
to four derivatives of the potential $V_M(R)$. These lead to contributions to the rovibrational energy (\ref{b10})
of $O\big((v+\thalf) (N(N+1))^2\big)$, which are obviously very small and are omitted in this paper for simplicity.

The new oscillation frequency $\w$ is then
\begin{equation}
\w \,=\, \w_0\,\Big[ 1 \,+\, \frac{1}{2}\Big(3 \,+\, \frac{R_0 V_M^{'''}(R_0)}{V_M^{''}(R_0)}\Big) \, \l^2\, N(N+1) 
\,+\, \ldots \,\Big] \ ,
\label{e5}
\end{equation}
with a corresponding contribution $(v+\thalf)\,\w$ to the energy. 

The term proportional to $V_M^{'''}(R_0)$ in (\ref{e5}), which contributes to the coefficient $\alpha_0$ below,
has an interesting interpretation in terms of ``vibrational stretching'' of the bond length in the presence of 
an asymmetric potential.  In particular, in the case of an anharmonic oscillator with a cubic interaction,
the expectation value of the position is shifted from the minimum of the potential.\footnote{Consider an 
anharmonic oscillator
with potential $V(x) = \thalf \mu \w_0^2 x^2 + g x^3$.  The expectation value $\langle \,x\,\rangle$ 
evaluated in the unperturbed SHO states gives 
\begin{equation*}
\langle \,x\,\rangle = - g \,\frac{3}{\mu^2 \omega_0^3} \, (v+\thalf) \ .
\end{equation*} 
Notice that the sign of $\langle x\rangle$ is opposite to the sign of the cubic anharmonic term in the potential.} 
Applying this result to the expansion of the angular momentum term $V_N(R)$ in (\ref{e1}),  
\begin{equation}
V_N(R) \,=\, \frac{1}{2} \l \,\w_0\,N(N+1)\,\Big( 1 \,-\, 2\,\frac{\langle \,x\,\rangle}{R_0} \,+\, \ldots  \Big) \ ,
\label{e6}
\end{equation}
where we write $R = R_0 + x$, and evaluating $\langle x\rangle$ approximating $V_M(R)$ as an 
anharmonic oscillator, we have
\begin{equation}
\frac{\langle\,x\,\rangle}{R_0} \,=\, -\frac{1}{2} \,\l \,\frac{R_0 V_M^{'''}(R_0)}{V_M^{''}(R_0)} \, (v + \thalf) \ ,
\label{e7}
\end{equation}
which reproduces the equivalent term in the energy found through the alternative route in (\ref{e5}).
This derivation makes clear that this is an increase in  \textit{kinetic} energy at fixed $N$ due to the 
reduction in mean bond length from (\ref{e7}).

Finally, collecting all these results, we find the coefficients in the expansion
\begin{align}
E_{v N} \,=\, &(v+\thalf) \,\w_0  \,-\,  x_0 \,(v+\thalf)^2 \,\w_0 \nonumber \\
&+\, B_0 \,N(N+1) \,\w_0   \,-\,  \a_0 \,(v + \thalf) N(N+1) \,\w_0 \,-\,  D_0 \,(N(N+1))^2 \,\w_0 \,+\,  \ldots 
\label{e8}
\end{align}
of the rovibrational energies are
\begin{align}
B_0 \,&=\, \frac{1}{2}\,\l    ~\ ,~~~~
&\alpha_0 \,&=\, -\frac{1}{2} \Big( 3 \,+\, \frac{R_0 V_M^{'''}}{V_M^{''}} \,\Big) \, \l^2   ~\ , \nonumber \\
D_0 \,&=\, \frac{1}{2} \,\l^3  ~\ ,  
&x_0 \,&=\, \left[\frac{5}{48}\, \Big(\frac{R_0 V_M^{'''}}{V_M^{''}}\Big)^2 \,-\, \frac{1}{16}\, 
\Big(\frac{R_0^2 V_M^{(4)}}{V_M^{''}}\Big)\,\right]\, \l  ~\ ,
\label{e9}
\end{align}
where all the indicated derivatives of $V_M(R)$ are taken at $R_0$.  

The result for $x_0$ arises from taking into account the anharmonic nature of the potential
and including the $O(v+\thalf)^2$ contributions to the energy of an anharmonic oscillator with
a $gx^3 \,+\, hx^4$ potential, where we identify $g = \tfrac{1}{6} V_M^{'''}$ and $h = \tfrac{1}{24} V_M^{(4)}$.
These are proportional to $h$ at first order and $g^2$ at second order. 
The explicit formulae, which determine $x_0$ in (\ref{e9}), are quoted in footnote \ref{foot AHO}.

Importantly, each quoted coefficient is the leading term in an expansion in $\l$.
At each order, the prefactors are known numbers involving successively higher derivatives of the 
basic inter-nucleon potential.  We have derived these in Appendix \ref{appendix A}
to a precision sufficient for our purpose here, which is to determine the leading corrections 
to (\ref{e8}) when Lorentz and \textsf{CPT} violating interactions are introduced.

\vskip0.5cm
We now need to extend this calculation to include the Lorentz and \textsf{CPT} violating potential
$V_{\textrm{SME}}^e(R)$ in the nucleon Schr\"odinger equation.  The idea is to iterate the effective potential
method for the hierarchy of perturbations, first including only the angular momentum term as above,
then including the potential $V_{\textrm{SME}}^e(R)$ as a further small perturbation. As always, the
Lorentz and $\textsf{CPT}$ perturbations are considered the smallest, and we work only to first
order in the SME couplings so terms of $O(V_{\textrm{SME}}^e)^2$ are immediately dropped.

So here we define 
\begin{equation}
U_{\it eff}(R) \,=\, V_{\it eff}(R) \,+\, V_{\textrm{SME}}^e(R)  \ ,
\label{e10}
\end{equation}
and expand about the minimum $R_u$ of $U_{\it eff}(R)$. Setting $U_{\it eff}'(R_u) = 0$ with $R_u = R_m + \d R_u$,
we have
\begin{equation}
\frac{\d R_u}{R_m} \,=\, - \frac{1}{R_m V_{\it eff}^{''}(R_m)} \,V_{\textrm{SME}}^{e\,'}(R_m)\ .
\label{e11}
\end{equation}
We then find, repeating the steps (\ref{e3}) and (\ref{e4}) above and neglecting terms of $O(\textrm{SME})^2$, 
that 
\begin{align}
U_{\it eff}(R_u) \,&=\, U_{\it eff}(R_m) \,+\, \d R_u \, U_{\it eff}'(R_m)  \nonumber \\
&=\, V_{\it eff}(R_m) \,+\, V_{\textrm{SME}}^e(R_m) \ ,
\label{e12}
\end{align}
and 
\begin{equation}
U_{\it eff}^{''}(R_u) \,=\, V_{\it eff}^{''}(R_m)  \,+\, V_{\textrm{SME}}^{e\,''}(R_m) \,-\, 
\frac{V_{\it eff}^{'''}(R_m)}{V_{\it eff}^{''}(R_m)} \, V_{\textrm{SME}}^{e\,'}(R_m) \ ,
\label{e13}
\end{equation}
the latter giving the new effective vibration frequency $\w_u$.

The next step is to rewrite (\ref{e12}) and (\ref{e13}) in terms of derivatives at the minimum $R_0$ of the 
original potential $V_M(R)$, rather than at $R_m$. We have already found $V_{\it eff}(R_m)$ and 
$V_{\it eff}^{''}(R_m) $ in (\ref{e3}) and (\ref{e4}),
so we just need
\begin{align}
U_{\it eff}(R_u) \,&=\, V_{\it eff}(R_m) \,+\, V_{\textrm{SME}}^e(R_0) \,+\, \d R\, V_{\textrm{SME}}^{e\,'}(R_0)  \,+\,
\frac{1}{2} \d R^2 \, V_{\textrm{SME}}^{e\,''}(R_0)  \nonumber \\
&=\, V_{\it eff}(R_m) \,+\, V_{\textrm{SME}}^e(R_0)  \nonumber \\
&~~~~+\, \Big[\l^2\,N(N+1) - 
\Big( 3 \,+\, \half\frac{R_0 V_M^{'''}}{V_M^{''}} \,\Big) \,\l^4\,(N(N+1))^2 \Big] \, R_0\, V_{\textrm{SME}}^{e\,'}(R_0) \nonumber \\
&~~~~+\,\frac{1}{2} \l^4 \,(N(N+1))^2 \, R_0^2\, V_{\textrm{SME}}^{e\,''}(R_0) \ ,
\label{e14}
\end{align}
which gives the Lorentz and \textsf{CPT} violating contributions $B_{\rm SME}^e$ and $D_{\rm SME}^e$ in 
(\ref{e17}) below, then from (\ref{e13}), and after some lengthy but straightforward algebra, we find
\begin{align} 
U_{\it eff}^{''}(R_u) \,&=\, V_{\it eff}^{''}(R_m) \,+\,  V_{\textrm{SME}}^{e\,''}(R_0)\,-\, 
\frac{V_M^{'''}}{V_M^{''}}\, V_{\textrm{SME}}^{e\,'}(R_0)  \nonumber \\
&~~~+\, \l^2\,N(N+1) \Big[R_0\,V_{\textrm{SME}}^{e\,'''}(R_0) 
\,-\,\frac{V_M^{'''}}{V_M^{''}} \Big(R_0\, V_{\textrm{SME}}^{e\,''}(R_0)\,+\, \b\, V_{\textrm{SME}}^{e\,'}(R_0) \Big) \,\Big] \ ,
\label{e15}
\end{align}
again with the understanding that all derivatives of $V_M$ are taken at $R_0$. Here,
\begin{equation}
\b \,=\, 3 \,-\, 12\, \frac{V_M^{''}}{R_0\,V_M^{'''}} \,-\, \frac{R_0\,V_M^{'''}}{V_M^{''}}\, +\,
\frac{R_0\,V_M^{(4)}}{V_M^{'''}} \ .
\label{e16}
\end{equation}
The new effective vibration frequency is now given by $\w_u^2 = \w_0^2 \,U_{\it eff}^{''}(R_0)/ V_M^{''}(R_0)$
and the corresponding energy levels by $(v+\thalf)\,\w_u$.  This gives contributions to $\alpha_{\rm SME}^e$
and $\d_{\rm SME}^e$, defined below.

The Lorentz and \textsf{CPT} corrections to the rovibrational energies (\ref{e8}) can therefore be written 
in the form 
\begin{align}
E_{v NM_N} \,=\,\, V_{\rm SME}^e~ &+(1 + \d_{\rm SME}^e)\,(v+\thalf) \,\w_0  \,-\,  (x_0 + x_{\rm SME}^e)\,(v+\thalf)^2 \,\w_0 
\nonumber \\[5pt]
&+\, (B_0 + B_{\rm SME}^e) \,N(N+1) \,\w_0   \,-\,  (\a_0 +\a_{\rm SME}^e)\,(v + \thalf) N(N+1) \,\w_0 
\nonumber \\[5pt]
&-\,  (D_0 + D_{\rm SME}^e) \,(N(N+1))^2 \,\w_0 \,+\,  \ldots 
\label{e17}
\end{align}
with $V_{\rm SME}^e \equiv V_{\rm SME}^e(R_0)$ and coefficients,
\begin{align}
\d_{\rm SME}^e \,&=\, \frac{1}{2} \,\frac{1}{V_M^{''}}\, \Big[\, V_{\rm SME}^{e\,''}  \,-\, 
\frac{V_M^{'''}}{V_M^{''}}\,\, V_{\rm SME}^{e\,'}\Big] 
\nonumber \\[5pt]
B_{\rm SME}^e \,&=\,  \l\, \,\frac{1}{V_M^{''}}\, \Big[\frac{1}{R_0}\,V_{\rm SME}^{e\,'} \,\Big]  \nonumber  \\[5pt]
\alpha_{\rm SME}^e \,&=\, \l^2\, \,\frac{1}{V_M^{''}}\,\bigg[ -\frac{1}{2} R_0\, V_{\rm SME}^{e\,'''} \,+\, 
\frac{3}{4} \Big(1 + \frac{R_0\, V_M^{'''}}{V_M^{''}}\Big) \, V_{\rm SME}^{e\,''}  \nonumber \\
\,&~~~~~~~~~~~~~~~~~~~~~~~~~~~~
-\, \Big(6 \,+\, \frac{9}{4}\,\frac{R_0\, V_M^{'''}}{V_M^{''}} \,+\, \frac{3}{4} \Big(\frac{R_0\, V_M^{'''}}{V_M^{''}}\Big)^2 
\,-\, \frac{1}{2} \frac{R_0^2 \,V_M^{(4)}}{V_M^{''}} \Big) \, \frac{1}{R_0}\,V_{\rm SME}^{e\,'} \bigg]
\nonumber \\[5pt]
D_{\rm SME}^e \,&=\, \l^3 \, \,\frac{1}{V_M^{''}}\, \Big[-\frac{1}{2}  \,  V_{\rm SME}^{e\,''}  \,+\, 
\Big( 3 \,+\, \half\frac{R_0\, V_M^{'''}}{V_M^{''}} \,\Big) \,\frac{1}{R_0}\, V_{\rm SME}^{e\,' }  \,\Big] \ .
\label{e18}
\end{align}
Notice that the $\w_0$-independent term $V_{\rm SME}^e$ should be retained here since it depends
on the quantum numbers $N, M_N$ through (\ref{d23}) and so contributes to rovibrational transition
energies in which $\D N \neq 0$.

The final coefficient $x_{\rm SME}^e$ is found from the anharmonic terms as before, where here
$g = \tfrac{1}{6} U_{\it eff}^{'''}(R_u)$ and $h = \tfrac{1}{24}U_{\it eff}^{(4)}(R_u)$.  Since we are not calculating 
terms involving the angular momentum at $O(v+\thalf)^2$, we may simplify $U_{\it eff}(R)$ in (\ref{e10})
to be just $V_M(R) + V_{\rm SME}^{e}(R)$, with the corresponding simplification of $\d R_u$ in (\ref{e11}).
Then we use the AHO energies in footnote\footnote{Explicitly, for an anharmonic oscillator
with potential $V= \thalf \mu \w_0^2 x^2 \,+\, g x^3 \,+\, h x^4$, and working to 2nd order perturbation
theory in $g$, the kinetic energy $K = p^2/2\mu$ is
\begin{align*}
\langle\,K\,\rangle \,&=\, \frac{1}{2}(v+\thalf)\w_0 \,+\, \frac{3}{2} \frac{h}{\mu^2 \w_0^3}\, 
\big((v+\thalf)^2 +\tfrac{1}{4}\big) \w_0 \,
-\, \frac{1}{16}\frac{g^2}{\m^3 \w_0^5}\big(60(v + \thalf)^2 \,+\,7\big) \w_0  \nonumber \\
&=\,  \frac{1}{2} E_v^{(0)}  \,+\, E_v^{(1)}(\l) \,+\, E_v^{(2)}(g^2) \ ,
\end{align*}
where $E_v$ are the total energies at the indicated order.  \label{foot AHO}} 
to extract $x_{\rm SME}^e$, keeping terms of first
order in the SME couplings only.  There are two sources of extra terms beyond the simple extension
of $x_0$ in (\ref{e9}).  The first comes from writing the derivatives at $R_0$ rather than $R_u$; thus,
for example, $U_{\it eff}^{(4)}(R_u) \rta V_M^{(4)}(R_0)\,+\,V_{\rm SME}^{e\, (4)}(R_0)§\,+\, \d R_u \,V_M^{(5)}(R_0) $, 
which produces extra terms proportional to $V_{\rm SME}^{e\, '}$.  Next, we must remember that the frequency
in the AHO energies is now $\w_u$ not $\w_0$, where $\mu \,\w_u^2 = U_{\it eff}^{''}(R_u)$ in (\ref{e15}).
Including both corrections eventually gives
\begin{align}
x_{\rm SME}^e  \,&=\, \l \, \,\frac{1}{V_M^{''}}\,\bigg[\frac{5}{24}\,
R_0^2\, \frac{V_M^{'''}}{V_M^{''}} 
\,\bigg(\, V_{\rm SME}^{e\,'''} \,-\,  \frac{V_M^{'''}}{V_M^{''}}\, V_{\rm SME}^{e\,''} \,+\, 
\Big( \, \Big(\frac{V_M^{'''} }{V_M^{''}}\Big)^2    \,-\,\frac{V_M^{(4)}}{V_M^{''}}\,\Big)  V_{\rm SME}^{e\,'}\bigg) 
~~~~~~~~~~~~~~\nonumber \\[7pt]
\,&~~~~~~~~~~
-\, \frac{1}{16}\, R_0^2\,\bigg(\,  V_{\rm SME}^{e\,(4)} \,-\, \frac{V_M^{(4)}}{V_M^{''}}\, V_{\rm SME}^{e\,''} \,+\,
\Big(\,\frac{V_M^{'''} V_M^{(4)}}{(V_M^{''})^2}\,-\, \frac{V_M^{(5)}}{V_M^{''}}\,\Big)\, V_{\rm SME}^{e\,'} \,
\bigg)\bigg] \ .
\label{e19}
\end{align}

 We should perhaps emphasise that while these expressions appear complicated, the coefficients
of $V_{\rm SME}^{e}$ and its derivatives are just simple numerical factors determined by $V_M(R)$ and have 
been evaluated in Appendix \ref{appendix A}.  That said, {\it all} the $V_M$-dependent terms shown in these
coefficients are necessary, since there is no reason to discard those with higher derivatives of $V_M$
as being smaller.  As a check that the effective potential method used here has indeed correctly and
completely identified all the terms occurring at the required order in $\l$, we present an alternative,
especially systematic, perturbative method of evaluation in Appendix \ref{appendix B}.

\vskip0.3cm
At this point, in addition to the rovibrational energies themselves, we can deduce expressions for the
Lorentz and \textsf{CPT} violating effects on the mean inter-nucleon bond length and the dissociation energy.

As well as the $N$-dependent centrifugal stretching, the mean bond length is also affected by the
SME couplings. This contribution, $\Delta R_{SME}$, is simply identified in the effective potential
method as $\d R_u$ in (\ref{e11}). Expanding the potentials $V_{\rm SME}^{e\, '}(R_m)$ and $V_{\it eff}^{''}(R_m)$
about $R_0$ instead of $R_m$, we find
\begin{equation}
\Delta R_{\rm SME} \,=\, - \frac{1}{V_M^{''}}\,\Big[V_{\rm SME}^{e\, '} \,+\, \l^2\, N(N+1)\,
\Big(R_0\, V_{\rm SME}^{e\, ''} \,-\, \Big(3 \,+\, \frac{R_0 V_M^{'''}}{V_M^{''}}\Big)\, V_{\rm SME}^{e\, '}\,\Big) \,\Big] \ ,
\label{ee1}
\end{equation}
up to higher order terms of $O\big(\l^2\, N(N+1)\big)^2$.

We define the dissociation energy $E^{\it diss}$ as the difference between the energy of the ground state
and the value of the inter-nucleon potential in the large-$R$ limit. 
The SME correction to the dissociation energy is then,
\begin{equation}
\Delta\,E_{\rm SME}^{\it diss} \,=\, V_{\rm SME}^{e}(\infty) \,-\, V_{\rm SME}^{e}(R_0) \,-\, \frac{1}{2}\,\d_{\rm SME}^e\,\w_0 \ .
\label{ee2}
\end{equation}
This is evaluated explicitly in the following section, noting that since
$\langle p^a\,p^b \rangle \,\rta\, (1/3)\d^{ab}$ as $R\rta \infty$, reflecting the spherical symmetry 
in this limit, the second term in (\ref{d23}) for $V_{\rm SME}^{e}(\infty)$ vanishes.

\vskip0.5cm

The remaining contribution to the rovibrational energies comes from the direct contribution $\D E_{\rm SME}^n$
from the Lorentz and \textsf{CPT} violating proton couplings $E_{ij}^p$ in the nucleon Schr\"odinger equation.
Recall from (\ref{c8}) that this requires us to evaluate the expectation value
\begin{equation}
\Delta E_{\textrm{SME}}^n \,=\, \frac{2}{m_p^2} E_{ij}^p \,\langle v\, N M_N |\, P^i P^j \,|v \,N M_N\rangle \ ,
\label{e20}
\end{equation}
in the original nucleon states $|v \,N M_N\rangle$, with all quantities expressed in the \textsf{EXP} frame.
The analysis mirrors that described in detail in section \ref{sect 4}. First, we  expand $P^i P^j$ in spherical 
harmonics,
\begin{equation}
P_i\,P_j \,=\, P^2 \,\Big( \frac{1}{3} \,\d_{ij} \,+\, \sum_M C_{ij}^M\,Y_{2M}(\theta, \phi) \,\Big) \ ,
\label{e21}
\end{equation}
where $(\theta, \phi)$ are the spherical polar angles made by the molecular axis (and therefore $\boldsymbol{P}$)
in the \textsf{EXP} frame, {\it i.e.}  $P_i = |\boldsymbol{P}|(\sin\theta\cos\phi, \sin\theta\sin\phi, \cos\theta)$.
Noting then that $P_i P_j = P^2\, (R^T)_{iz} R_{zj}$, which of course is inherent in the construction of (\ref{d15}),
the coefficient $C_{ij}^M$ here is identical to that found in section \ref{sect 4}. Taking the expectation value
as in (\ref{d20}) and evaluating the Clebsch-Gordan equations (\ref{d22}), we find without further calculation that
\begin{equation}
\D E_{SME}^n \,=\, \langle v N|\,P^2\,|v N\rangle\, \frac{2}{m_p^2} \,\Big[\, \frac{1}{3}\,\tr\,E_{ij}^p \,-\,
\frac{1}{3} \,\, c_{N M_N}\,\, \tr_Y\,E_{ij}^p\,\Big]
\label{e22}
\end{equation}

The problem therefore reduces to finding the expectation value $\langle\,P^2\,\rangle$ of the nucleon 
momentum. In this case, we do not have a simple explicit wave function solving the Schr\"odinger equation, 
so a direct evaluation is not straightforward. However, since $P^2/2\mu$ is just the kinetic energy,
all we need in practice is to identify the kinetic energy part of the total rovibrational energy $E_{vN}$
in (\ref{e8}). This requires examining each of the coefficients $x_0, B_0,\alpha_0$ and $D_0$ derived 
above and deducing on physical grounds what is their {\it kinetic} energy component. 

First, write (\ref{e22}) as
\begin{equation}
\D E_{\rm SME}^n \,=\, \langle v\,N|\,K\,|v\,N\rangle \, \tilde{V}_{\rm SME}^n  \ ,
\label{e23}
\end{equation}
with
\begin{equation}
\tilde{V}_{\rm SME}^n \,=\,  \frac{2}{3} \frac{1}{m_p} \Big[\, \tr\,E_{ij}^p \,-\,
\,c_{N M_N}\,\, \tr_Y\,E_{ij}^p\,\Big] \ .
\label{e24}
\end{equation}
Then, adding this contribution to (\ref{e17}), the rovibrational energies including all the Lorentz
and \textsf{CPT} breaking contributions are written as
\begin{align}
E_{v NM_N} \,=\, V_{\rm SME}^e \, &+(1 + \d_{\rm SME}^e + \d_{\rm SME}^n)\,(v+\thalf) \,\w_0  \,
-\,  (x_0 + x_{\rm SME}^e + x_{\rm SME}^n)\,(v+\thalf)^2 \,\w_0 \nonumber \\[2pt]
&+\, (B_0 + B_{\rm SME}^e + B_{\rm SME}^n) \,N(N+1) \,\w_0   \,  \nonumber \\[2pt]
&-\,  (\a_0 +\a_{\rm SME}^e + \a_{\rm SME}^n)\,(v + \thalf) N(N+1) \,\w_0 \nonumber \\[2pt]
&-\,  (D_0 + D_{\rm SME}^e + D_{\rm SME}^n) \,(N(N+1))^2 \,\w_0 \,+\,  \ldots 
\label{e25}
\end{align}

Beginning with the angular momentum independent terms, $x_0$ is found to $O(\l)$ by approximating 
$V_M(R)$ as an anharmonic oscillator with cubic and quartic interactions.  An explicit calculation,
consistent with the virial theorem, shows that at $O(v+1/2)$ the kinetic energy is $E_v/2$ while
at $O(v+1/2)^2$ the whole energy is kinetic (see footnote \ref{foot AHO}).
It follows immediately that the corresponding coefficients in (\ref{e25}) are
\begin{equation}
\d_{\rm SME}^n \,=\, \frac{1}{2}\, \tilde{V}_{\rm SME}^n \ , ~~~~~~~~~~~~~~~~
x_{\rm SME}^n \,=\, x_0\, \tilde{V}_{\rm SME}^n  \ .~~~~~~~~~~~
\label{e26}
\end{equation}

Turning to the $N$-dependent terms, it is clear that $B_0$ is a pure kinetic energy term.
However, For $D_0$ we emphasised above there were two contributions, weighted -1 and 1/2,
which arose due to centrifugal stretching. Of these the first represents a reduction in kinetic
energy, while the second is potential energy due to stretching the mean bond length.
This means we must take
\begin{equation}
B_{\rm SME}^n \,=\, B_0 \,\tilde{V}_{\rm SME}^n \ , ~~~~~~~~~~~~~~
D_{\rm SME}^n \,=\, 2 \,D_0\, \tilde{V}_{\rm SME}^n  \ . ~~~~~~~
\label{e27}
\end{equation}
For $\a_{\rm SME}^n$, an interpretation based on the effective potential is less clear, but a careful analysis 
using the methods in Appendix \ref{appendix B} shows that it follows the same pattern as above,
and we have
\begin{equation}
\a_{\rm SME}^n \,=\, \frac{3}{2}\, \a_0 \, \tilde{V}_{\rm SME}^n  \ . 
 ~~~~~~~~~~~~~~~~~~~~~~~~~~~~~~~~~~~~~~~~~~~~~~~~
\label{e28}
\end{equation}

Together, (\ref{e24}) to (\ref{e28}) complete the derivation of the Lorentz and \textsf{CPT} violating contributions 
to the rovibrational energy levels.

\vskip1.8cm

\section{The rovibrational spectrum  }\label{sect 6}

In this section, we draw together all the results of sections \ref{sect 4} and \ref{sect 5} and the appendices
to describe the implications of Lorentz and \textsf{CPT} violation for the rovibrational spectrum
of the (anti-)hydrogen molecular ion. 

First, we verify that our approximate methods describe the rovibrational spectrum in the absence
of Lorentz and \textsf{CPT} symmetry breaking sufficiently for our objective, that is to constrain
the SME couplings with greater precision than is possible with atomic (anti-)hydrogen alone.  
For this, we need the numerical results for $V_M(R)$ and its derivatives tabulated in Appendix
\ref{appendix A}, Table \ref{Table1}.  The key result is for the vibrational angular frequency $\w_0$.
Converting from the atomic units of Table \ref{Table1} using $\hat{R}_H = 1/(2 \hat{\mu}\, \hat{a}_0^2)$, 
we find (with $\hbar = c= 1$),
\begin{equation}
\w_0 \,=\, \sqrt{\frac{2 V_M^{''}(R_0)}{m_p}} ~=~ 0.020\, \hat{R}_H ~=~ 0.275\,{\rm eV} \ ,
\label{f1}
\end{equation}
which in terms of the fundamental QED parameters implies $w_0 \sim R_H \sqrt{\frac{m_e}{m_p}}$.
In standard spectroscopic units (with $h = c =1$), $ 1\, {\rm eV} \,=\,
2.418 \times 10^{14}\, {\rm Hz} \,=\, 8065.5\, {\rm cm}^{-1} $, so the corresponding frequency
is $2218\, {\rm cm}^{-1}$.

The dimensionless expansion parameter $\l$ is then determined for the $H_2^+$ ion as 
$\l \,=\, 1/(\mu \w_0 R_0^2) \,=\, 0.027$.  Note that this is parametrically $\l \sim \sqrt{\frac{m_e}{m_p}}$.

The coefficients of the expansion (\ref{e8}) of the rovibrational energy levels,
\begin{align}
E_{v N} \,=\, &(v+\thalf) \,\w_0  \,-\,  x_0 \,(v+\thalf)^2 \,\w_0 \nonumber \\
&+\, B_0 \,N(N+1) \,\w_0   \,-\,  \a_0 \,(v + \thalf) N(N+1) \,\w_0 \,-\,  D_0 \,(N(N+1))^2 \,\w_0 \,+\,  \ldots 
\label{f2}
\end{align}
are then found from Table \ref{Table1} to be
\begin{align}
B_0 \,&=\,\frac{1}{2}\,\l \,=\,  0.0135 \ ,   ~~~~~~~~~~~~~~~~~~
\a_0\,=\, 1.133\,\l^2 \,=\, 0.819\times 10^{-3} \ ,   \nonumber\\[3pt]
D_0\,&=\, \frac{1}{2}\,\l^3 \,=\, 1.944 \times 10^{-5}   ~~~~~~~~~~~~
x_0\,=\, 1.230\,\l \,=\,0.033 \,\ .
\label{f3}
\end{align}
Substituting back, we find the rovibrational levels in spectroscopic units of $\rm{cm}^{-1}$ 
in the form,
\begin{align}
E_{v N} \,=\, &2218\, (v+\thalf)  \,-\,  73.2 \,(v+\thalf)^2 \, \nonumber \\
&+\,29.82 \,N(N+1)    \,-\, 1.82 \,(v + \thalf) N(N+1)  \,-\, 0.0216 \,(N(N+1))^2  \,+\,  \ldots 
\label{f4}
\end{align}
which may be compared with precision calculations and data.\footnote{As a reference, we compare with the
corresponding result quoted in \cite{Varshalovich} (see also \cite{Korobov:2017}),
\begin{align*}
E_{v N} \,=\, & 2322.99\, (v+\thalf)  \,-\, 67.361  \,(v+\thalf)^2 \, \nonumber \\
&+\, 29.944 \,N(N+1)    \,-\,1.591   \,(v + \thalf) N(N+1) \, \,-\, 0.0198 \,(N(N+1))^2 \, \,+\,  \ldots 
%\label{Varsalovicheq}
\end{align*}
Comparing with (\ref{f4}), we find agreement at better than 1\% for $B_0$, less than $5\%$ for $\w_0$,
and within around 10\% for the others. This is reasonable given the very simplistic model of the
electron wavefunction used in Appendix \ref{appendix A}, and will be quite sufficient for their r\^ole below 
in determing the prefactors of the SME couplings.
} 
The hierarchy of the coefficients in powers of $\l$ is immediately evident.
A particular point of interest already here is that the term in (\ref{e9}) involving the third derivative $V_M^{'''}$
is essential even to give the correct sign for the $O(v+\thalf)\, N(N+1)$ contribution.
Moreover, the $x_0$ coefficient, which is determined entirely by the higher derivative terms in $V_M(R)$,
has the correct value and sign to produce the narrowing of vibrational energy level splittings as
$v$ becomes larger, allowing roughly 20 vibrational states below the dissociation energy. 
Both these observations confirm the necessity of including a full analysis of the anharmonic terms
in $V_M(R)$ to produce a realistic match to the rovibrational spectrum and ensure our results remain
valid beyond the lowest vibrational states.

We now turn to the Lorentz and \textsf{CPT} violating effects in the rovibrational spectrum.  Writing the 
coefficients $\d _{\rm SME}^e$, $B_{\rm SME}^e, \ldots$ in (\ref{e18}), (\ref{e19}) in terms of $V_{\rm SME}^{e}$ 
and its derivatives using the numerical values in Table \ref{Table1} gives, in atomic units,
\begin{align}
\d_{\rm SME}^e \,&=\,\, \big[\,2.669\, V_{\rm SME}^{e\, ''}  \,+\,  7.018 \, V_{\rm SME}^{e\, '} \,\big]   \nonumber  \\[10pt]
B_{\rm SME}^e \,&=\,  \,\,\l\, \big[\, 2.664 \, V_{\rm SME}^{e\, ' } \,\big]  \nonumber \\[10pt]
\a_{\rm SME}^e \,&=\, \l^2\,\big[-5.346\,V_{\rm SME}^{e\, '''} \,-\, 17.084\, V_{\rm SME}^{e\, '' } \,-\, 
3.408\, V_{\rm SME}^{e\, '} \,\big]   \nonumber \\[10pt]
D_{\rm SME}^e \,&=\, \l^3\,\big[ \, -2.669 \,V_{\rm SME}^{e\, ''} \,+\, 0.897\, V_{\rm SME}^{e\, '}\, \big]
\nonumber \\[10pt]
x_{\rm SME}^e \,&=\, \,\,\l\,\big[\, -1.337\, V_{\rm SME}^{e\, (4)} \,-\, 11.738\, V_{\rm SME}^{e\, '''}
\,-\, 22.004\, V_{\rm SME}^{e\, ''} \,-\, 5.215\, V_{\rm SME}^{e\, '}\, \big]  \ .
\label{f5}
\end{align}

The next step is to re-express $V_{\rm SME}^{e }$ in (\ref{d23}) in atomic units, as assumed here.
Recalling $R_H \,=\, 1/(2 \, m_e\,a_0^2)$,\footnote{From now on we ignore the distinction between
$\hat{\mu},\, \hat{a}_0,\,\hat{R}_H$ and $m_e,\,a_0,\,R_H$ which is of course numerically very small.}
we write $V_{\rm SME}^{e}(R)$
as
\begin{equation}
V_{\rm SME}^{e}(R)\,=\, \Big[\,\frac{2}{3}\, \tr\,\langle p^a\,p^b\rangle\, \,\frac{1}{m_e}\,\tr\, \tilde{E}_{ij}^{e} ~+~
\frac{1}{3} \,\tr_Y\langle p^a\,p^b\rangle\,\, c_{NM_N}\, \frac{1}{m_e}\,\tr_Y\tilde{E}_{ij}^{e} \,\Big]\ ,
\label{f6}
\end{equation}
where recall the shorthand,
\begin{equation}
c_{NM_N} \,=\, \frac{N(N+1) - 3 M_N^2}{(2N-1)(2N+3) } \ ,
\label{f7}
\end{equation}
and the momentum expectation values and their derivatives take their numerical values as given
in Table \ref{Table1}. 
Substituting these values, we can express the coefficients in (\ref{f5}) directly in terms of the SME
couplings $\tr\, \tilde{E}_{ij}^{e}$ and $\tr_Y\tilde{E}_{ij}^{e}$ as follows:
\begin{align}
V_{\rm SME}^e \,&=\, \,\big[ \,0.782\, \frac{1}{m_e}\,\tr\, \tilde{E}_{ij}^{e}   
\,+\, 0.120\, c_{NM_N}\, \frac{1}{m_e}\,\tr_Y \tilde{E}_{ij}^{e} \,\big]   \nonumber  \\[10pt]
\d_{\rm SME}^e \,&=\, \,\big[  -1.000\,   \frac{1}{m_e}\,\tr\, \tilde{E}_{ij}^{e}   
\,-\, 0.272\, c_{NM_N}\, \frac{1}{m_e}\,\tr_Y \tilde{E}_{ij}^{e} \,\big]   \nonumber  \\[10pt]
B_{\rm SME}^e \,&=\,\,\, \l\,\big[ -0.666 \, \frac{1}{m_e}\,\tr\, \tilde{E}_{ij}^{e}  
\,-\,0.112 \, c_{NM_N}\, \frac{1}{m_e}\,\tr_Y \tilde{E}_{ij}^{e}  \,\big]  \nonumber  \\[10pt]
\a_{\rm SME}^e \,&=\,   \l^2\,\big[-2.152\,  \frac{1}{m_e}\,\tr\, \tilde{E}_{ij}^{e}  
\,-\, 0.095\, c_{NM_N}\, \frac{1}{m_e}\,\tr_Y  \tilde{E}_{ij}^{e}\,\big]  \nonumber  \\[10pt]%
D_{\rm SME}^e \,&=\, \l^3\,\big[  -0.979\,   \frac{1}{m_e}\,\tr\, \tilde{E}_{ij}^{e} 
\,-\, 0.060\, c_{NM_N}\, \frac{1}{m_e}\,\tr_Y  \tilde{E}_{ij}^{e}  \,\big]  \nonumber  \\[10pt]
x_{\rm SME}^e \,&=\, \,\, \l\, \big[  -1.629\, \frac{1}{m_e}\,\tr\, \tilde{E}_{ij}^{e}   
\,-\, 0.065\,\, c_{NM_N}\, \frac{1}{m_e}\,\tr_Y \tilde{E}_{ij}^{e} \,\big]  \ .
\label{f8}
\end{align}

\newpage

For reference, we also quote here the equivalent results for the coefficients $\d_{\rm SME}^n, \, B_{\rm SME}^n,
\ldots$ in the same format.  From (\ref{e24}), (\ref{e26})-(\ref{e28}) and (\ref{f3}) we find:
\begin{align}
\d_{\rm SME}^n \,&=\,\,\, 0.333\,\big[\, \frac{1}{m_p}\,\tr\, {E}_{ij}^{p}   
\,-\, c_{NM_N}\, \frac{1}{m_p}\,\tr_Y {E}_{ij}^{p} \,\big]   \nonumber  \\[10pt]
B_{\rm SME}^n \,&=\,\,0.333 \,\l\,\big[\, \frac{1}{m_p}\,\tr\, {E}_{ij}^{p}  
\,-\, c_{NM_N}\, \frac{1}{m_p}\,\tr_Y {E}_{ij}^{p}  \,\big]  \nonumber  \\[10pt]
\a_{\rm SME}^n \,&=\,   1.133\,\l^2\,\big[\,  \frac{1}{m_p}\,\tr\, {E}_{ij}^{p}  
\,-\, c_{NM_N}\, \frac{1}{m_p}\,\tr_Y {E}_{ij}^{p}\,\big]  \nonumber  \\[10pt]%
D_{\rm SME}^n \,&=\, 0.667\,\l^3\,\big[\,   \frac{1}{m_p}\,\tr\, {E}_{ij}^{p} 
\,-\, c_{NM_N}\, \frac{1}{m_p}\,\tr_Y {E}_{ij}^{p}  \,\big]  \nonumber  \\[10pt]
x_{\rm SME}^n \,&=\, \,0.820\, \l\, \big[\, \frac{1}{m_p}\,\tr\, {E}_{ij}^{p}   
\,-\,\, c_{NM_N}\, \frac{1}{m_p}\,\tr_Y {E}_{ij}^{p} \,\big]  \ .
\label{f9}
\end{align}

We can also write the expressions for the SME corrections to the mean bond length
and dissociation energy in this form.  From the leading term in (\ref{ee1}), omitting the
weak $N$-dependence of $O\big(\l^2\,N(N+1)\big)$ here, we find
\begin{equation}
\Delta R_{\rm SME} \,=\, \big[\, 1.340\,   \frac{1}{m_e}\,\tr\, \tilde{E}_{ij}^{e}   
\,+\, 0.226\, c_{NM_N}\, \frac{1}{m_e}\,\tr_Y \tilde{E}_{ij}^{e} \,\big] \ .
\label{ff1}
\end{equation}
The dissociation energy is defined in (\ref{ee2}) in terms of $V_{\rm SME}^{e}(R_0)$, given  
in (\ref{f6}), $V_{\rm SME}^{e}(\infty)$, where $\tr\,\langle p^a\, p^b\rangle\,\rta\, 1$
and $\tr_Y\langle\ p^a\,p^b\rangle \,\rta\, 0$, and $\d_{\rm SME}^e$, which modifies the ground state
energy. Substituting from Table \ref{Table1}, we then find
\begin{equation}
\Delta E_{\rm SME}^{\it diss} \,=\, \big[  -0.105\, \frac{1}{m_e}\,\tr\, \tilde{E}_{ij}^{e}   
\,-\, 0.117\,\, c_{NM_N}\, \frac{1}{m_e}\,\tr_Y \tilde{E}_{ij}^{e} \,\big]  \ ,
\label{ff2}
\end{equation}
where both $\Delta R_{\rm SME}$ and $\Delta E_{\rm SME}^{\it diss}$ in
(\ref{ff1}) and (\ref{ff2}) are in atomic units.

\vskip0.5cm
Finally, to make contact with previous work, we should express $\tr\,\tilde{E}_{ij}^{e},\, 
\tr_Y\tilde{E}_{ij}^{e}$ and $\tr\,{E}_{ij}^{p},\, \tr_Y {E}_{ij}^{p}$ directly in terms 
of the original SME couplings in the Lagrangian (\ref{a1}).

It is often convenient in spectroscopy applications to use a spherical harmonic decomposition
of the SME couplings.  
To make the translation, we need to compare the 
non-relativistic Hamiltonian $H_{\rm SME}^{\rm NR}$ in (\ref{c1}) with its equivalent in terms of the
SME couplings $c_{njm}^{\rm NR}$ and $a_{njm}^{\rm NR}$ in the spherical harmonic basis,\footnote{Note
the rather unintuitive but now standard notation (see {\it e.g.}~\cite{Kostelecky:2013rta})  
where mass terms are inserted in these definitions
so that both $c_{njm}^{\rm NR}$ and $a_{njm}^{\rm NR}$ have the {\it same} dimensions, {\it i.e.}
mass dimension $1-n$, unlike the corresponding Lagrangian couplings $c_{\m\n}$ and 
$a_{\m\n\l}$, where $c_{\m\n}$ is dimensionless.}
\begin{equation}
H_{\rm SME}^{\rm NR} \,=\, -\sum_{njm}\, \big(c_{njm}^{\rm NR}\,-\, a_{njm}^{\rm NR}\big)\, |\boldsymbol{p}|^n\,
Y_{jm}(\hat{\boldsymbol{p}}) \,+\, \ldots
\label{f10}
\end{equation}
where $\ldots$ represents the spin-dependent couplings in $B_k$,  $D_k$ and $F_{ijk}$.
Specialising to the terms with $n=2$, we then write
\begin{equation}
p^i\,p^j \,=\, \Big(\frac{1}{3} \d_{ij} \,+\, \sum_m\, C_{ij}^m\, Y_{2m}(\hat{\boldsymbol{p}})\,\Big)
\, |\boldsymbol{p}|^2 \ ,
\label{f11}
\end{equation}
and comparing coefficients of the spherical harmonics between (\ref{c1}) and (\ref{f10}),
using (\ref{d19}) for $C_{ij}^0$, we identify
\begin{equation}
\frac{1}{m}\,\tr\, {E}_{ij} \,=\, -3m\, \frac{1}{\sqrt{4\pi}}\,\big(c_{200}^{\rm NR} \,-\, a_{200}^{\rm NR}\big) \ , 
~~~~~~~~~~~~
\frac{1}{m}\,\tr_Y {E}_{ij} \,=\, 3m\, \sqrt{\frac{5}{4\pi}}\,\big(c_{220}^{\rm NR} \,-\, a_{220}^{\rm NR}\big)  \ .
~~~~\\[3pt]
\label{f12}
\end{equation}
with the obvious extension to $\tilde{E}_{ij}^e = E_{ij}^e \,+\,\frac{1}{2}(m_e^2/m_p^2) E_{ij}^p$,\,
and $E_{ij}^p$.
In terms of the original SME couplings in (\ref{c2}), we therefore have 
\begin{align}
\frac{1}{\sqrt{4\pi}}\,c_{200}^{\rm NR} \,&=\,\,\frac{1}{3m}\big(c_{ii} \,+\, \tfrac{3}{2}c_{00}\big) 
~=~\frac{5}{6m}\,c_{00}    \ ,
~~~~~~~~~
\frac{1}{\sqrt{4\pi}}\,a_{200}^{\rm NR} \,=\,\, a_{0ii}  \,+\, a_{000} \ , \nonumber \\[5pt]
\sqrt{\frac{5}{4\pi}}\,c_{220}^{\rm NR} \,&=\, -\frac{1}{3m}\, \tr_Y c_{ij} \ ,
~~~~~~~~~~~~~~~~~~~~~~~~~~~~~~
\sqrt{\frac{5}{4\pi}}\,a_{220}^{\rm NR} \,=\, - \tr_Ya_{0ij} \ ,
\label{f13}
\end{align}
recalling that $c_{\m\n}$ should be chosen to have vanishing spacetime trace \cite{Colladay:1998fq}.

Altogether, equations (\ref{e25}) with (\ref{f3}), (\ref{f8}), (\ref{f9}) and (\ref{f12}) complete our 
description of the dependence of the rovibrational energy levels $E_{vN M_N}$ of the molecular 
(anti-)hydrogen ion on the spin-independent SME couplings.

\section{Rovibrational transitions}\label{sect 7} 

In this final section, we take a first look at how the results in section \ref{sect 6} may be used to constrain
the Lorentz and \textsf{CPT} couplings through measurements of the rovibrational transitions 
${\rm H}_2^{+}$, and eventually its antimatter counterpart $\overline{\rm H}_2^{\,-}$, 
and explain why these offer the possibility of improving existing bounds on the \textsf{CPT} odd spin-independent
SME couplings by several orders of magnitude.
Further discussion, together with an assessment of current and future experimental possibilities, 
will be presented elsewhere.

We begin by describing some basic features of the spectrum of the molecular hydrogen ion
(see, for example, \cite{SchillerCP}).
Since the $1s\,\sigma_g$ electron state in ${\rm H}_2^{+}$ is symmetric, the total nucleon state 
must be antisymmetric under exchange of the two protons -- this implies that for {\it even} $N$,
the two protons are in an antisymmetric spin state $I=0$ (where $\boldsymbol{I}$ is the sum of the spins 
of the two protons), while for {\it odd} $N$, the spin state is symmetric so $I=1$. 
These two groupings are referred to as Para- and Ortho-${\rm H}_2^+$ respectively.
Since ${\rm H}_2^{+}$ is a homonuclear molecule, it has no permanent dipole moment
and electric dipole transitions (denoted E1) between rovibrational states (which would allow
$\D N = 1$ transitions) are forbidden.
The remaining possibilities are single-photon electric quadrupole (E2) and two-photon (TP) 
transitions, for which the selection rule $\D N \,=\, 0,\, \pm 2$ applies (with $N=0\rta N=0$ 
transitions disallowed for E2 but permitted for TP only).  These selection rules therefore 
forbid transitions between the Ortho and Para states, so their spectra are essentially independent
of each other. 
%This is sketched in Fig.~\ref{FigSpectrum}.

To illustrate some of the possibilities of constraining the SME couplings by precision measurements 
of rovibrational transitions, we consider here just the leading coefficients $\d_{\rm SME}^{e,\,p}$ and 
$B_{\rm SME}^{e,\, p}$ from (\ref{f9}). In this case,    
\begin{align}
&E_{v N M_N} \,=\, \Big[- \frac{\,\,\,1}{\sqrt{4\pi}}\Big(
2.35 \,m_e \big(\tilde{c}_{200}^{{\rm NR}\,e}
\,-\, \tilde{a}_{200}^{{\rm NR}\,e} \big) \Big)
\,+\, \sqrt{\frac{5}{4\pi}}\,c_{NM_N}\Big(
0.36 \,m_e \big(\tilde{c}_{220}^{{\rm NR}\,e}
\,-\, \tilde{a}_{220}^{{\rm NR}\,e} \big) \Big)\,\Big]
\nonumber \\[5pt]
&~~~~~~~~~~~~~~ +\, (v+\thalf) \,\w_0\,\Big[ \,1\,+\, \frac{\,\,\,1}{\sqrt{4\pi}} \Big(
3.00\, m_e \big(\tilde{c}_{200}^{{\rm NR}\,e}
\,-\, \tilde{a}_{200}^{{\rm NR}\,e} \big) 
\,-\,m_p \big(c_{200}^{{\rm NR}\,p} \,-\,a_{200}^{{\rm NR}\,p} \big)  \Big) \nonumber \\[5pt]
\,&~~~~~~~~~~~~~~~~~~~~~~~~~~~
+\, \sqrt{\frac{5}{4\pi}} \, c_{NM_N}\, \Big(- 0.82 \,m_e\big(\tilde{c}_{220}^{{\rm NR}\, e} \,-\, 
\tilde{a}_{220}^{{\rm NR}\, e}\big) \,-\, 
m_p\big(c_{220}^{{\rm NR}\,p} - a_{220}^{{\rm NR}\,p}\big) \,\Big) \,\Big]
\nonumber \\[7pt]
&~~~~~ + \, \frac{1}{2} \l\, N(N+1)\, \w_0\, \Big[ \,1\,+\, 
\frac{\,\,\,1}{\sqrt{4\pi}} \Big( 3.99\,m_e\big(\tilde{c}_{200}^{{\rm NR}\, e}\,-\, \tilde{a}_{200}^{{\rm NR}\, e} \big)
\,-\, 2\,m_p\big(c_{200}^{{\rm NR}\,p} \,-\,a_{200}^{{\rm NR}\,p} \big) \Big) \nonumber \\[5pt]
\,&~~~~~~~~~~~~~~~~~~~~~~~~~
+\, \sqrt{\frac{5}{4\pi}} \, c_{NM_N}\, \Big(- 0.67\,m_e\big(\tilde{c}_{220}^{{\rm NR}\, e} \,-\, 
\tilde{a}_{220}^{{\rm NR}\, e}\big) \,-\, 
2\,m_p\big(c_{220}^{{\rm NR}\,p} - a_{220}^{{\rm NR}\,p}\big) \,\Big) \,\Big] \ ,
\label{g1}
\end{align}
with $\tilde{c}_{200}^{{\rm NR}\, e} = c_{200}^{{\rm NR}\, e} + 
\tfrac{1}{2} \,c_{200}^{{\rm NR}\,p}$, {\it etc.}

The simplest case is a transition where the angular momentum quantum number is unchanged,
{\it i.e.}  $\D v \neq0$ but $\D N = 0$ (and $\D M_N = 0$).  In this case,
\begin{align}
\frac{\D E_{\rm SME}}{\D E} \,&=\, \d_{\rm SME}^e \,+\, \d_{\rm SME}^p  \nonumber \\
&=\, \frac{\,\,\,1}{\sqrt{4\pi}}\,\Big[\, 3.00\,m_e\big(\tilde{c}_{200}^{{\rm NR}\, e}\,-\, 
\tilde{a}_{200}^{{\rm NR}\, e} \big)
\,-\, 1.00\,m_p\big(c_{200}^{{\rm NR}\,p} \,-\,a_{200}^{{\rm NR}\,p} \big)  \nonumber \\[5pt]
\,&~~
+\, c_{N M_N}\,\sqrt{5}\,
\, \Big(- 0.82\,m_e\big(\tilde{c}_{220}^{{\rm NR}\, e} \,-\, \tilde{a}_{220}^{{\rm NR}\, e}\big) \,-\, 
m_p\big(c_{220}^{{\rm NR}\,p} - a_{220}^{{\rm NR}\,p}\big) \,\Big) \,\Big] \ .
\label{g2}
\end{align}
This already demonstrates a number of important features. First, even for a purely vibrational 
transition, the Clebsch-Gordan factor $c_{N M_N}$ arising with the $(njm)\,=\, (220)$ 
SME couplings implies that the transition frequency is dependent on $N$ and $M_N$.  
By comparing transitions with different $N$, it is then possible to separate the contributions
from the $(200)$ and $(220)$ couplings, which can therefore be constrained individually.

Conversely, it is evident from (\ref{g1}) that even a purely rotational transition \cite{Alighanbari:2020}
with $\D N\neq 0$ but $\D v = 0$ will also depend on the vibrational quantum number $v$
through the $(220)$ couplings.

Furthermore, because of the different numerical coefficients of the electron and proton
terms in $\d_{\rm SME}$ and $B_{\rm SME}$, and $V_{\rm SME}^e$,
by comparing different transitions where $\D v \neq 0,\, \D N=0$, 
or $\D v=0, \,\D N \neq 0$, or both non-zero,  it is also possible to distinguish  the contributions from  
$\big(\tilde{c}_{200}^{{\rm NR}\, e}\,-\, \tilde{a}_{200}^{{\rm NR}\, e} \big)$ and the purely proton coupling
$\big(c_{200}^{{\rm NR}\, p}\,-\, a_{200}^{{\rm NR}\, p} \big)$. Similarly for the $(220)$ couplings.
A relatively small number of rovibrational transitions in ${\rm H}_2^+$ could therefore lead
to precision constraints on all four types of $(c_{njm}^{\rm NR}\,-\, a_{njm}^{\rm NR})$ couplings
in (\ref{g1}).

Finally, if similar precision experiments become possible with the antihydrogen molecular
ion \cite{Myers, Karr:2018, Zammit,Zammit:2025nky},
then because the $c_{njm}^{\rm NR}$ couplings are \textsf{CPT} even 
while the $a_{njm}^{\rm NR}$ couplings are \textsf{CPT}-odd, 
taking the difference of the ${\rm H}_2^{+}$ and $\overline{\rm H}_2^{\, -}$
spectra would isolate the dependence on the $a_{njm}^{\rm NR}$ couplings. 
This would allow individual constraints to be placed on all 8 SME couplings in (\ref{g1}).

\vskip0.3cm
To see why these constraints are potentially so powerful for the molecular ions, we should compare
with the equivalent transitions with atomic hydrogen and antihydrogen. 
Keeping only the spin-independent SME couplings, we find for the $1{S}_d$\,-\,$2{S}_d$
transition \cite{Kostelecky:2015nma, Charlton:2020kie} measured for antihydrogen 
by ALPHA \cite{Ahmadi:2018eca}
(where the suffix $d$ labels the particular hyperfine state),
\begin{equation}
\frac{\D E^{\rm SME}_{1S_d\,-\,2S_d}}{\D E_{1S\,-\,2S} }\,=\, \frac{2 m_e}{\sqrt{4\pi}}\Big[
\big(c_{200}^{{\rm NR}\, e}\,-\, a_{200}^{{\rm NR}\, e} \big) \,+\, 
\big(c_{200}^{{\rm NR}\, p}\,-\, a_{200}^{{\rm NR}\, p} \big) \,\Big] \ .
\label{g3}
\end{equation}
Notice this is only sensitive to the $(njm) = (200)$ couplings. 

To access a transition sensitive to the $(njm)=(220)$ couplings in (anti-)hydrogen, we need to 
consider a state with electron orbital momentum $\ell \neq 0$ (in the same way that we require 
$N\neq0$ to access the $(220)$ contributions to the rovibrational states in (\ref{g1})).  
ALPHA has measured the $1S_d$\,-\,$2P_{c-}$ transition \cite{ALPHA:2020rbx} 
for which we can show \cite{Kostelecky:2015nma, Charlton:2020kie},
\begin{align}
\frac{\D E^{\rm SME}_{1S_d\,-\,2P_{c-}}}{\D E_{1S\,-\,2S} } \,&=\,
\frac{2 m_e}{\sqrt{4\pi}}\, \Big[\big(c_{200}^{{\rm NR}\, e}\,-\, a_{200}^{{\rm NR}\, e} \big) \,+\, 
\big(c_{200}^{{\rm NR}\, p}\,-\, a_{200}^{{\rm NR}\, p} \big) \nonumber \\[3pt]
&~~~~~~~~~ \,-\, \frac{\sqrt{5}}{\,\,30}\,(1\,+\,3\cos{ 2\eta})\,\Big(\,
\big({c}_{220}^{{\rm NR}\, e} \,-\, {a}_{220}^{{\rm NR}\, e}\big) \,+\, 
\big(c_{220}^{{\rm NR}\,p} - a_{220}^{{\rm NR}\,p}\big) \,\Big)\, \Big] \ ,  
\label{g4}
\end{align}
where $\cos\eta$ is a magnetic field dependent mixing angle.

This brings us to the key point. Focusing on the \textsf{CPT} violating couplings,
the two-photon $1S$\,-$2S$ transition in (anti-)hydrogen constrains the combination
\begin{equation}
m_e \,\big(a_{200}^{{\rm NR}\, e} \,+\,a_{200}^{{\rm NR}\, p}\big) \,\lesssim\, R_{1S-2S}  \ ,
\label{g5}
\end{equation}
where $R_{1S-2S}$ is the relative precision of the measurement. ALPHA has measured
this transition for antihydrogen with a precision $R_{1S-2S}\simeq 2 \times 10^{-12} $ 
\cite{Ahmadi:2018eca}, while a precision of $O(10^{-15})$ is known for hydrogen 
\cite{Parthey:2011lfa, Matveev:2013orb}.
Knowing this, the single-photon $1S$\,-$2P$ transition, which is measured by ALPHA
with the much weaker precision $R_{1S-2P} \simeq 10^{-8}$ \cite{ALPHA:2020rbx}, 
constrains the $(220)$ combination,
\begin{equation}
m_e \,\big(a_{220}^{{\rm NR}\, e} \,+\,a_{220}^{{\rm NR}\, p}\big) \,\lesssim\, R_{1S-2P}  \ .
\label{g6}
\end{equation}
With just four measurements, $1S$\,-\,$2S$ and $1S$\,-\,$2P$ for hydrogen and antihydrogen,
we are able to constrain only the combinations (\ref{g5}), (\ref{g6}) of electron and proton
couplings, and similar for the $c_{njm}^{\rm NR}$.  Crucially, {\it both} the electron and proton
couplings are multiplied by $m_e$.

In contrast, for ${\rm H}_2^+$, by measuring only a small number of rovibrational transitions
as indicated above, together with the equivalent for $\overline{\rm H}_2^{\,-}$, 
we have sufficient data to constrain all 8 coefficients.
Denoting the precisions generically by $R_{rovib}$, in this case we find the constraints,
\begin{equation}
m_e \,\big(a_{200}^{{\rm NR}\, e}\,+\,\tfrac{1}{2} a_{200}^{{\rm NR}\, p} \big) \,\lesssim\, R_{rovib}\ , 
~~~~~~~~~~~~~
m_p \,a_{200}^{{\rm NR}\, p} \,\lesssim\, R_{rovib}\ , 
\label{g7}
\end{equation}
and
\begin{equation}
m_e \,\big(a_{220}^{{\rm NR}\, e} \,+\, \tfrac{1}{2} a_{200}^{{\rm NR}\, p} \big)\lesssim\, R_{rovib}\ , 
~~~~~~~~~~~~~~
m_p \,a_{220}^{{\rm NR}\, p} \,\lesssim\, R_{rovib}\ ,
\label{g8}
\end{equation}
with equivalent bounds for the $c_{njm}^{\rm NR}$.
Here, we have written out in full the couplings $\tilde{c}_{njm}^{\rm NR\, e}$ and 
$\tilde{a}_{njm}^{\rm NR\, e}$.  Notice that the bounds on the proton SME couplings arising from
their indirect influence on the electron Schr\"odinger equation have the same $m_e$ mass 
dependence as those found in the atomic transitions. This is not true of the proton couplings 
arising directly in the nucleon Schr\"odinger equation, which involve $m_p$ itself.
  
So not only do the rovibrational transitions allow us to constrain the electron and proton
couplings separately, the constraint on the proton couplings $a_{200}^{{\rm NR}\, p}$ and 
$a_{220}^{{\rm NR}\, p}$ is {\it enhanced} by a factor $m_p/m_e \sim 10^3$ compared to the 
atomic transitions.  This is in addition to the potential enhancement from the greater
precision of frequency measurements for the rovibrational transitions, which of course is 
especially marked for the $(220)$ couplings where the precision of the $1S$\,-\,$2P$
measurement in antihydrogen is necessarily 4 orders of magnitude below that 
achieved for $1S$\,-\,$2S$.

In practice, by far the most stringent bounds on the proton $(c^{{\rm NR}\,p}_{2jm} - a^{{\rm NR}\,p}_{2jm})$
couplings arise not from atomic hydrogen transitions but from those associated
with heavier elements, in particular ${^{133}}Cs$ clock transitions
\cite{Kostelecky:1999mr,Wolf:2006uu,Pihan-LeBars:2016pjg}. 
However, unlike the molecular ions discussed here, these do not have accessible 
antimatter counterparts and cannot isolate the \textsf{CPT} odd couplings. From the standpoint
of the SME, therefore, the unique strength of spectroscopy of the (anti-)molecular ions ${\rm H}_2^+$
and $\overline{\rm H}_2^-$ is in providing a direct test of \textsf{CPT} symmetry at
high precision.

\vskip0.5cm

So far, we have only considered the rovibrational states to be described by quantum numbers
$v, N, M_N$ and have neglected spin (see \cite{Korobov2006, KKH20081, KKH20082}). 
In general, the $1s\s_g$ states in ${\rm H}_2^+$  will be 
labelled as $|v\, N\, I\, S\, F\, J\, M_J\rangle$, according to the angular momentum addition 
scheme $\boldsymbol{F} = \boldsymbol{I} + \boldsymbol{S}$, where $\boldsymbol{I}$ and 
$\boldsymbol{S}$ are the total nucleon and electron spins respectively, 
followed by $\boldsymbol{J} = \boldsymbol{N} + \boldsymbol{F}$.

The hyperfine structure is then determined by a Hamiltonian incorporating 5 combinations
of spin-spin and spin-orbit interactions, with known coefficients calculated at $O(\alpha^2)$
to six-figure precision in QED \cite{Korobov2006}. 
This simplifies greatly for the case of Para-$\textrm{H}_2^{\,+}$ 
and we illustrate our results for this case here. The energy eigenstates, at zero background
magnetic field, are then $|v\, N\, S\,  J\, M_J\rangle$ with $S=1/2$ always and 
$\boldsymbol{J} = \boldsymbol{N} + \boldsymbol{S}$.  The corresponding hyperfine 
interaction is simply,
\begin{equation}
H_{\rm HFS} \,=\, c_e(v,N)\, \boldsymbol{N}.\boldsymbol{S} ~~=~  \frac{1}{2} c_e(v,N)\,
\big( \boldsymbol{J}^2 - \boldsymbol{N}^2 - \boldsymbol{S}^2 \big) \ .
\label{g9}
\end{equation}
The values of $c_e(v,N)$ for low values of $v$ and $N$ are given in Table 1 of \cite{Korobov2006}.
It follows directly that the rovibrational energies are split by the value of $J = N\pm 1/2$
as follows,
\begin{align}
\D E_{\rm HFS} \,&=\,\, \frac{1}{2}\,c_e(v,N)\, N \ , ~~~~~ (J = N+\thalf) \nonumber \\
&=\, -\frac{1}{2}\, c_e(v,N) \,(N+1)  \ , ~~~~~ (J = N-\thalf) \ ,
\label{g10}
\end{align}
and are degenerate with respect to $M_J$.  

To calculate the SME contributions to the hyperfine states, we first express the eigenstates
$|J\,M_J\rangle$ in terms of the $|M_N\, M_S\rangle$ basis states through Clebsch-Gordan
coefficients as follows,
\begin{equation}
|J\,M_J\rangle\,\,=\,\, \sum_{M_S} \, C^{J\,M_J}_{N\,M_N,\,\tfrac{1}{2}\,M_S}\,\, |M_N \, M_S\rangle \ ,
~~~~~~~(M_N = M_J - M_S) \ .
\label{g11}
\end{equation}
Now, at the point (\ref{d16}) for the electron $V^e_{\rm SME}(R)$, or (\ref{e22}) for the proton
$\tilde{V}^n_{\rm SME}$, our derivation of the rovibrational energies requires the evaluation 
of $\sum_M\, C_{ij}^M\, Y_{2M}$ between eigenstates, taken there to be $|v\,N\,M_N\rangle$.
Using the hyperfine states $|J\,M_J\rangle$ instead, we find
\begin{align}
&E_{ij} \, \sum_M\, C_{ij}^M\, \langle J\,M_J|Y_{2M}\,|J\,M_J\rangle \nonumber \\
&~~~~~~~~~~~= \,E_{ij}\,\, \sum_{M_S} \, C_{ij}^0\, \big(C^{J\,M_J}_{N\,M_N,\,\tfrac{1}{2}\,M_S}\big)^2 \,
\langle M_N\,M_S|\,Y_{20}\,|M_N\,M_S\rangle ~~~~~~~ \nonumber \\
&~~~~~~~~~~~\,= \,-\frac{1}{3}\, \tr_Y E_{ij}\, \,\sum_{M_S}\, 
\big(C^{J\,M_J}_{N\,M_N,\,\tfrac{1}{2}\,M_S}\big)^2 \,\, c_{N(M_J-M_S)} \ ,
\label{g12}
\end{align}
in terms of the factor $c_{N M_N}$ in (\ref{f7}). The results of these calculations are presented 
in Paper 2 \cite{Shore:2025zor}, 
showing in detail how the spin-independent SME couplings break the degeneracy 
with respect to $M_J$ of the hyperfine energy levels. 

\vskip0.3cm
In practice, spectroscopy will be performed in a background magnetic field, which we 
may use to define the $3$-axis of the \textsf{EXP} frame, $\boldsymbol{B}= (0,0,B)$.
The rovibrational energies will therefore acquire a shift due to the Zeeman effect.
Restricting again to Para-$\textrm{H}_2^{\,+}$, the Zeeman Hamiltonian is
\begin{equation}
H_Z \,=\, g_e\m_B\,\boldsymbol{S}\cdot\boldsymbol{B} \,-\,  
g_m(v,N) \m_B\,\boldsymbol{N}\cdot\boldsymbol{B} \ ,
\label{g15}
\end{equation}
where the effective $g$-factor $g_m(v,N)$ depends on the rovibrational quantum 
numbers \cite{KKH20081,Shore:2025zor}. It is $O(m_e/m_p)$ so the electron spin
dominates the energy shifts. For large $B$, where the hyperfine splitting is negligible
compared to the Zeeman effect, the combined hyperfine-Zeeman eigenstates become
approximately just the $|v\,N\,S\,M_N\,M_S\rangle$ states.
For intermediate magnetic fields, apart from the extremal states 
$|J=N+\tfrac{1}{2},M_J=\pm(N+\tfrac{1}{2})\rangle \,\equiv\, 
|M_N=\pm N,\,M_S=\pm\tfrac{1}{2}\rangle$
where the two bases coincide, the remaining $4N$ hyperfine-Zeeman eigenstates are mixed
pairwise, with $2N$ $B$-dependent mixing angles.\footnote{For an explicit example, the
hyperfine-Zeeman states in Para-$\textrm{H}_2^{\,+}$ with $N=2$ and the associated
energy curves as a function of $B$ interpolating between pure hyperfine and 
large-$B$ Zeeman states, see sections C, D and Fig.~2 of the Supplementary Material of
\cite{SchillerCP} and Paper 2 \cite{Shore:2025zor}.}

Our original derivation of the rovibrational energy levels therefore applies directly in
the large $B$ regime where the eigenstates are essentially given by $|M_N \, M_S\rangle$,
and the results above with $c_{NM_N}$ hold, showing the extra $M_N$ dependence of the
energy levels due to the SME couplings. For intermediate values of the magnetic field,
the corresponding SME energies involving $c_{220}^{\rm NR}$ and $a_{220}^{\rm NR}$ will 
depend in an individual way on the state mixing, described in detail 
in Paper 2. 

\vskip0.3cm
Our focus in this paper has been on the spin-independent SME couplings, 
for reasons given above. 
Inclusion of the spin-dependent couplings is in principle a straightforward extension of 
the results given here, though inevitably more complicated in general given the 
spin dependence of the hyperfine states. Again, our results are fully described in 
Paper 2. 

\vskip0.3cm
Finally, apart from precision measurements of rovibrational energy levels, including a 
comparison of ${\rm H}_2^{\,+}$ and $\overline{\rm H}_2^{\,-}$ as a direct \textsf{CPT} test, 
Lorentz symmetry breaking would reveal itself through annual (and sidereal) variations 
in the transition frequencies resulting from the SME coupling contributions derived above. 
This analysis is well-known (see {\it e.g.}~\cite{Kostelecky:2008ts}) and we only 
comment briefly here. 

The first step is to rewrite the SME couplings in components referred to the standard 
\textsf{SUN} frame (see section \ref{sect 3}). In general this involves a rotation from the
\textsf{EXP} to \textsf{SUN} frames, which depends on the location and orientation of
the background magnetic field of the particular experiment. Note, however, that this
rotation does not affect the isotropic $(200)$ couplings, which retain their form in the
\textsf{SUN} frame.

These \textsf{SUN} frame couplings are then subject to a Lorentz transformation
with velocity $v_{\oplus}$ of $O(10^{-4})$ corresponding to the Earth's rotation around the Sun,
resulting in small periodic variations with the Earth's orbit. In \textsf{SUN} frame coordinates
$(T,X,Y,Z)$, the orbital velocity is 
\begin{equation}
\boldsymbol{v}_{\oplus}\,=\,v_{\oplus} \,\Big( \sin\Omega_{\oplus}T\, \boldsymbol{e}_X
-\cos\Omega_{\oplus}T \big(\cos\eta \,\boldsymbol{e}_Y \,+\,
\sin\eta \,\boldsymbol{e}_Z \big) \Big) \ ,
\label{g18}
\end{equation}
where $\Omega_{\oplus}$ is the orbital frequency and $\eta = 23.4^{\circ}$ is the
tilt angle between the Earth's equator and the orbital plane.
The isotropic SME couplings in the rovibrational energy levels are simply
$\tfrac{1}{\sqrt{4\pi}} \,c_{200}^{\rm NR} \,=\, \tfrac{1}{3 m} \big(\tfrac{3}{2} c_{TT} + c_{KK}\big)
\,=\,\tfrac{5}{6m} c_{TT}$ 
and $\tfrac{1}{\sqrt{4\pi}} \,a_{200}^{\rm NR} \,=\, a_{TTT} + a_{TKK}$.
Applying a Lorentz transformation with $\boldsymbol{v}_{\oplus}$
then gives the periodic variation of the combinations of isotropic
SME couplings appearing in the rovibrational energies (\ref{g1}) as,
\begin{align}
\tfrac{1}{\sqrt{4\pi}}\, \d_{\oplus} \big(c_{200}^{\rm NR} - a_{200}^{\rm NR} \big) 
\,&=\, v_{\oplus} \sin\Omega_{\oplus}T \Big(\tfrac{5}{3m} c_{XT} - 5 a_{XTT} - a_{XKK}\Big)
\nonumber \\[3pt]
&~~~- v_{\oplus}\cos\Omega_{\oplus}T 
\Big( \cos\eta \, \big( \tfrac{5}{3m} c_{YT} - 5 a_{YTT} - a_{YKK}\big)  \nonumber \\
&~~~~~~~~~~~~~~~~~~~~~~~
+\sin\eta \, \big(\tfrac{5}{3m} c_{ZT}  - 5 a_{ZTT} - a_{ZKK} \big)\Big) \ .
\label{g19}
\end{align}
Similar results for the non-isotropic combination ($c_{220}^{\rm NR} - a_{220}^{\rm NR})$
depend in detail on the orientation of the \textsf{EXP} frame for a specific experiment.
Note that the variations introduce a dependence on different components of the SME
couplings from those appearing in the rovibrational energies themselves.
Detection of such annual variations of the ultra-precise rovibrational transition frequencies 
would be a clear signal of Lorentz, and potentially \textsf{CPT}, violation.

\vskip2cm
\noindent {\large{\textbf{Acknowledgements}}}
\vskip0.5cm
I am grateful to Stefan Eriksson for motivating this research 
and for many helpful discussions during its progress.  I would also like to thank the 
Higgs Centre for Theoretical Physics at the University of Edinburgh for hospitality 
in the course of this work and the CERN Theory Division for Visiting Scientist support.

\newpage

\appendix{

\section{Solution of the electron Schr\"odinger equation}\label{appendix A}

In this appendix, we give some details of the solution of the electron Schr\"odinger equation
in the absence of Lorentz and \textsf{CPT} violation.
In particular, we determine the energy eigenvalue $E_e(R)$, which carries through to the
nucleon Schr\"odinger equation in the Born-Oppenheimer analysis as the inter-nucleon potential $V_M(R)$.
We also calculate the $R$-dependence of the momentum expectation values $\langle p^a p^b\rangle$, 
which are needed as input in determining the potential $V_{\textrm{SME}}^e(R)$ in (\ref{c6}).

From the main text, the Schr\"odinger equation for the electron wavefunction $\psi(\boldsymbol{r};R)$ is
\begin{equation}
\left(- \frac{1}{2\hat{\mu}} \nabla_{\boldsymbol{r}}^2 \,+\, V_{mol}(R,r_{1e},r_{2e})  \right) \psi(\boldsymbol{r};R) 
\,=\, E_e(R) \,\psi(\boldsymbol{r};R) \ ,
\label{ap1}
\end{equation}
with the binding potential $V_{mol}(R,r_{1e},r_{2e})$ given in
(\ref{b2}).  The basic idea of the solution is to make an {\it ansatz} for the electron wavefunction in the 
$1s\sigma_g$ state comprising a sum of hydrogen-like $1s$  wave functions, {\it viz.}
\begin{equation}
\psi(\boldsymbol{r};R) \,=\, \frac{1}{\sqrt{2(1 + I_0(R))}} \left( \psi_H(r_{1e};R) \,+\, \psi_H(r_{2e};R)\right) \ ,
\label{ap2}
\end{equation}
where 
\begin{equation}
\psi_H(r;R) \,=\, \sqrt{\frac{\c(R)^3}{\pi \hat{a}_0^3}}\, e^{- \c(R) r/\hat{a}_0} \ .
\label{ap3}
\end{equation}
and the ``overlap function'' $I_0(R)$ is required to normalise the wavefunction. Explicitly,
\begin{equation}
I_0(R) \,=\, \int d^3\boldsymbol{r}\,\psi_H(r_{1e}) \,\psi_H(r_{2e}) \ .
\label{ap4}
\end{equation}

The necessity of including the scaling factor $\c(R)$ of the effective Bohr radius in (\ref{ap3}) is seen by considering
the limits of large and small nucleon separation.  As $R\rta \infty$, one of $r_{1e}$ or $r_{2e}$ in (\ref{ap3})
becomes large and so the total wavefunction reduces to a single $1s$ hydrogen state. The molecule has
dissociated leaving a hydrogen atom and an isolated proton. 
In this limit, therefore, we require $\c(R) \rta 1$, recovering the original
Bohr radius.  At the other extreme, as $R\rta 0$ the two nucleons coalesce and the wavefunction should reduce to 
that of a hydrogen-like atom with $Z=2$. In this case the dependence on the Bohr radius implies $\c(R) \rta 2$.  
We see below how this is realised in the explicit numerical solutions for the energies and momenta.

In (\ref{ap3}), we have introduced the notation $\hat{a}_0 = 1/\alpha \hat{\mu}$ for  the ``reduced Bohr radius'' 
appropriate to the reduced electron mass $\hat{\mu}$ in section \ref{sect 2}.  Together with the
corresponding ``reduced Rydberg constant'' $\hat{R}_H = \thalf \alpha^2 \hat{\mu} = 
 1/(2\hat{a}_0^2\, \hat{\mu})$, this defines the units of length and energy used from now on.
Rescaling to these ``atomic units'', the electron Schr\"odinger equation is simply
\begin{equation}
\left(- \nabla_{\boldsymbol{r}}^2 \,+\, 2 \Bigl(\frac{1}{R}\, -\, \frac{1}{r_{ie}}   \,-\, \frac{1}{r_{2e}}\Bigr) \right)
\psi(\boldsymbol{r};R) \,=\, E_e(R) \,\psi(\boldsymbol{r};R) \ ,
\label{ap5}
\end{equation}
where all quantities are now dimensionless. 

To calculate the energy eigenvalues $E_e(R)$, we need a number of elementary integrals, all of which can
be evaluated analytically for $\c(R) = 1$ \cite{QM}.  Defining the kinetic and potential energies
as $K(R)$ and $U(R)$ respectively, we have
\begin{equation}
K(R) \,=\, -\langle \psi|\,\nabla_{\boldsymbol{r}}^2 \,|\psi\rangle \ , ~~~~~~~~~~~~~~
U(R)\,=\, 2 \langle\psi| \,\frac{1}{R}\, -\, \frac{1}{r_{1e}}   \,-\, \frac{1}{r_{2e}} \,|\psi\rangle \ .
\label{ap6}
\end{equation}
First, for the overlap function we need 
\begin{equation}
I_0 \,=\, \frac{1}{\pi} \int d^3\boldsymbol{r} \, e^{-r_{1e}} \, e^{-r_{2e}} ~~=~~\bigl(1 + R + \frac{1}{3} R^2 \bigr) e^{-R} \ .
\label{ap7}
\end{equation}
Then, for the potential energy,
\begin{align}
I_1 \,&=\, \frac{1}{\pi} \int d^3\boldsymbol{r} \, \frac{1}{r_{1e}} \,e^{-r_{1e}} \, e^{-r_{1e}} ~~=~~
1  \ ,\\
I_2 \,&=\, \frac{1}{\pi} \int d^3\boldsymbol{r} \, \frac{1}{r_{1e}} \,e^{-r_{2e}} \, e^{-r_{2e}} ~~=~~
\frac{1}{R} \,-\, \frac{1}{R} (1 + R) e^{-2R} \ ,  \\
I_3 \,&=\, \frac{1}{\pi} \int d^3\boldsymbol{r} \, \frac{1}{r_{1e}} \,e^{-r_{1e}} \, e^{-r_{2e}} ~~=~~
(1+R)e^{-R}\ ,   
\label{ap8}
\end{align}
so that in total, with $\c(R)$ set to 1,
\begin{equation}
U_1(R) \,=\, \frac{2}{R} \,-\, \frac{2}{(1 + I_0)} ( I _1 + I_2 + 2I_3) \ .
\label{ap11}
\end{equation}
Here, and below, we use the suffix 1 to indicate that the associated quantity is evaluated
with $\c(R) \rightarrow 1$. 
For the kinetic energy, a similar calculation gives 
\begin{equation}
K_1(R)\,=\, -1 \,+\, \frac{2}{(1 + I_0)} ( I_1 + I_3 ) \ .
\label{ap12}
\end{equation}
Separating out the hydrogen $1s$ energy in reduced Rydberg units, $\hat{E}_{1s} = -1$, we may write
the total energy eigenvalue as
\begin{align}
E_{e1}(R) \,&=\, K_1(R) \,+\, U_1(R)  \nonumber \\
&=\, \hat{E}_{1s} \,+\, \frac{2}{R} \,-\, \frac{2}{(1 + I_0)} (I_2 + I_3) \ . 
\label{ap13}
\end{align}
This shows the typical Morse function form, with a minimum at $R_0 = 2.493$.

\vskip0.2cm
To find the Lorentz and \textsf{CPT} violating contributions $V_{\textrm{SME}}^e(\boldsymbol{R})$, we also need the
expectation values of the components of the momenta, specifically
\begin{equation}
\langle \,p^a \,p^b\,\rangle \,=\, -\int d^3\boldsymbol{r}\, \psi(\boldsymbol{r};R) \,\frac{\partial}{\partial x_a} 
\frac{\partial}{\partial x_b}\,
\psi(\boldsymbol{r};R) \ .
\label{ap14}
\end{equation}
The required integrals in this case include
\begin{align}
J_1 \,&=\, - \frac{1}{\pi} \int d^3\boldsymbol{r} \,e^{-r_{1e}} \, \frac{\partial^2}{\partial x^2} \, e^{-r_{1e}} 
~~=~~ \frac{1}{3} \ ,  \\
J_2 \,&=\, - \frac{1}{\pi} \int d^3\boldsymbol{r} \,e^{-r_{2e}} \, \frac{\partial^2}{\partial x^2} \, e^{-r_{2e}} 
~~=~~ \frac{1}{3} \ , \\
J_3 \,&=\, - \frac{1}{\pi} \int d^3\boldsymbol{r} \,e^{-r_{1e}} \, \frac{\partial^2}{\partial x^2} \, e^{-r_{2e}} 
~~=~~ \frac{1}{3} (1+R) e^{-R}\ .
\label{ap15}
\end{align}
Cylindrical symmetry in the \textsf{MOL} frame ensures identical results with derivatives 
$\partial^2/\partial y^2$, while for $\partial^2/\partial z^2$ we find 
$J_3 \rta \tfrac{1}{3}(1 + R -R^2) e^{-R}$ with  $J_1$ and $J_2$ unchanged.
It is also apparent from the evaluation that the mixed terms, $\langle p^a p^b\rangle$ with $a\ne b$,
vanish.

Collecting these results we find, again with $\c(R)$ set to 1,
\begin{equation}
\langle\,p^x p^x\,\rangle_1 \,=\, \langle p^y p^y \rangle_1 \,=\, 
\frac{1}{3} \frac{1}{(1 + I_0)}\, \bigl( 1 + (1+R) e^{-R}\bigr) \ ,
\label{ap18}
\end{equation}
and
\begin{equation}
\langle\,p^z p^z\,\rangle_1 \,=\, \frac{1}{3} \frac{1}{(1 + I_0)}\, \bigl( 1 + (1+R-R^2) e^{-R}\bigr) \ ,
\label{ap19}
\end{equation}
and as an immediate consistency check we verify
$K_1 = \langle p^x p^x\rangle_1  +   \langle p^y p^y\rangle_1 + \langle p^z p^z\rangle_1$.   

\vskip0.2cm
Now introduce the rescaling factor $\c(R)$ into the electron wavefunctions. By inspection, we see that
the kinetic and potential energies and the momentum expectation values now become 
\begin{equation}
K(R) \,=\, \c(R)^2 K_1(\c(R) R) \ ,  ~~~~~~~~~~~~~~~~
U(R) \,=\, \c(R) \,U_1(\c(R) R)  \ ,
\label{ap20}
\end{equation}
and 
\begin{equation}
\langle\, p^a p^b\,\rangle(R) \,=\, \c(R)^2 \,\langle\,p^a p^b\,\rangle_1(\c(R) R) \ .
\label{ap21}
\end{equation}
The aim here is to find a function $\c(R)$ which minimises the energy $E_e(R)$, where
\begin{equation}
E_e(R) \,=\, \c(R)^2 K_1(\c(R) R) \,+\, \c(R) \,U_1(\c(R) R)   \ .
\label{ap22}
\end{equation}
We use a numerical method\footnote{What follows is our interpretation of the method applied in
\cite{Muller:2004tc}, and we may compare our numerical results with the table of values for $E_e(R_0)$,
$\c_(R_0)$, $\langle p^a p^b\rangle(R_0)$ and their derivatives at $R_0$ quoted there. 
\vskip0.2cm
%For low derivatives the agreement is very good, with some divergence as the number of derivatives
%increases, presumably due to the different numerical techniques used.
}, the idea 
being to choose a discrete set of values $R=R_n$ and for each $n$ find the value of $\c(R_n)$ for which 
the energy $E_e(R_n)$ is minimised. We then construct an interpolating function $\c(R)$ to fit this
set of values. The result is the function $\c(R)$ shown in Fig.~\ref{Fig gamma}.
Notice that this does indeed satisfy the requirement that $\c(0) = 2$ and $\c(R) \rta 1$ for large $R$,
though this is not imposed as a constraint.
\begin{figure}[h!]
\centering{ {\includegraphics[scale=0.56]{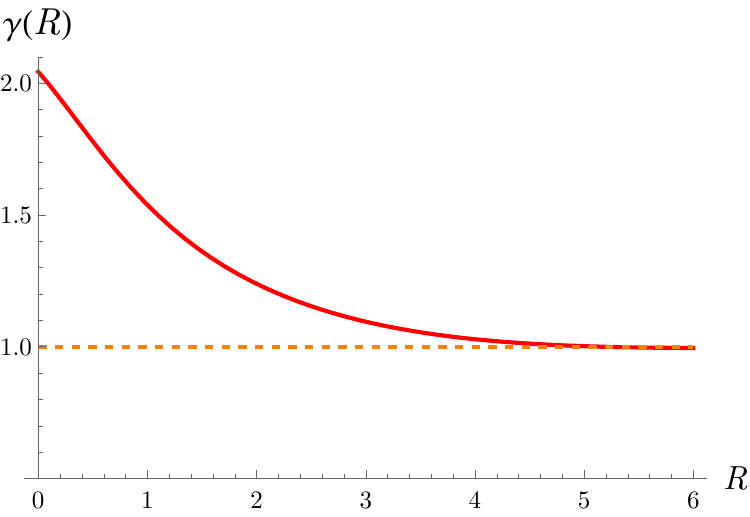} }  \hskip0.5cm
{\includegraphics[scale=0.58]{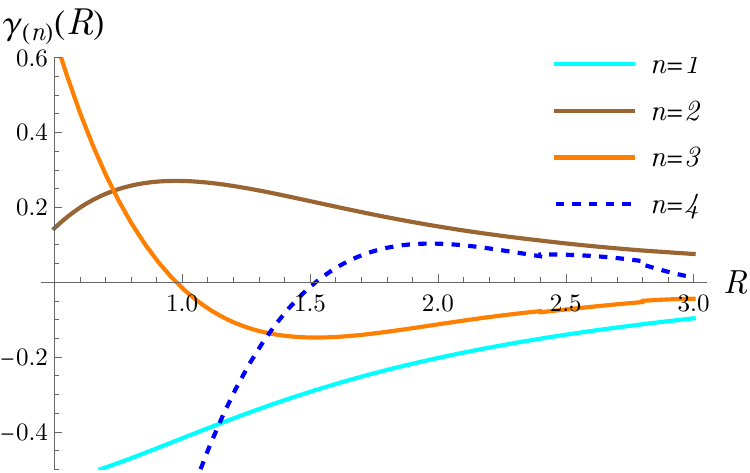}} } 
\caption{The interpolating function $\gamma(R)$ giving the appropriate scaling of the electron wave 
function as the inter-nucleon distance $R$ is varied.  The right hand figure shows its first four derivatives
in the vicinity of the minimum of the potential at $R_0 = 2.003$. See Table \ref{Table1}.  }
\label{Fig gamma}
\end{figure}

Given $\c(R)$, the energy and momentum expectation values are found from (\ref{ap22}) and (\ref{ap21}).
These are shown in Figs.~\ref{FigVM} and \ref{Figmomenta}, 
A few comments are worth making here. Note that the energy
$E_e(R)$ has a deeper minimum with the improved ansatz with the interpolating function $\c(R)$,
and it occurs at the smaller value $R_0 = 2.003$.  This gives a much improved agreement with the
measured dissociation energy for ${\rm H}_2^+$, though still not sufficiently deep. However, the value for the
fundamental vibration frequency $\omega_0 =\sqrt{ V_M^{''}(R_0)/\mu} = 0.020\,R_H$,
in good agreement with the measured value.
  
\begin{figure}[h!]
\centering{{\includegraphics[scale=0.57]{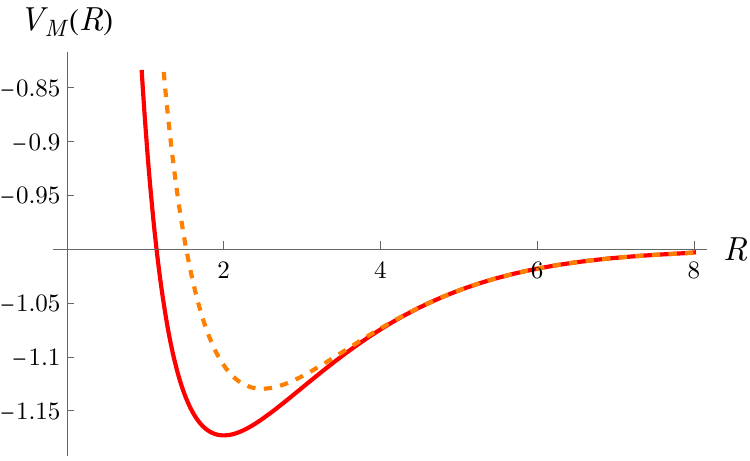}} \hskip 0.5cm 
{\includegraphics[scale=0.57]{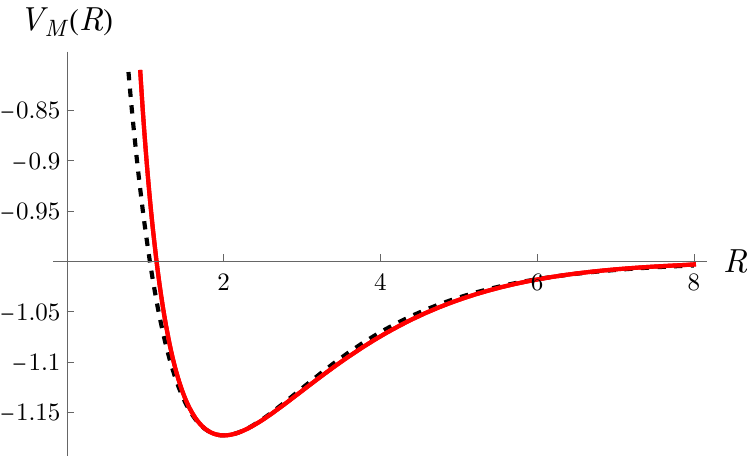}} }
\caption{The nucleon potential $V_M(R) \equiv E_e(R)$ as the inter-nucleon distance, or bond length,
 $R$ is varied. The minimum is at $R_0 = 2.003$. 
The left hand figure shows a comparison with the equivalent with the scaling factor $\gamma(R)$
left fixed equal to 1 (orange, dashed). The right hand figure shows the comparison with a 
Morse potential (black, dashed) with its two free parameters set to match the potential minimum 
$V_M(R_0$ and curvature $V_M^{''}(R_0)$.}
\label{FigVM}
\end{figure}

As expected, for large $R$ where the molecule has effectively separated, 
$E_e(R)$ tends to the atomic $1s$ level $\hat{E}_{1s} = -1$. Moreover, we numerically confirm 
the analytic results $K(R)\rta 1$ and $U(R)\rta -2$ as $R\rta \infty$, in agreement with the virial theorem
for a Coulomb interaction.\footnote{The virial theorem for a power-law potential $V(x)$ states
\begin{equation*}
2\langle\, K\, \rangle \,=\,  \langle\,x \frac{dV}{dx} \,\rangle \ ,
\end{equation*}
so for the Coulomb potential in \ref{ap5}, and neglecting the inter-nucleon repulsion which tends to
zero for large $R$, this implies $2\langle K\rangle = - \langle U \rangle$.\label{fn VT}} 
On the other hand, for $R\rta 0$, if we set aside the divergent nucleon-nucleon interaction term 
$2/R$ in the potential, we find with $\c(0) = 2$ that $K(0) \rta 4$ and $U(0) \rta -8$, again
satisfying the virial theorem and confirmed numerically in Fig.~\ref{FigVM}.  A comparison of 
$V_M(R)$ with the phenomenological Morse potential\footnote{The Morse potential is a 
two-parameter function,
\begin{equation*}
V_{Morse}(R) \,=\, D_e \big(1- e^{-a x}\big)^2 - D_e
\end{equation*} 
where $x= R - R_0$.  To make the comparison, we fit the parameters $D_e$ and $a$ to the value 
of $V_M(R)$ and its second derivative at $R_0$, {\it viz.}~$D_e = - V_M(R_0)$ and 
$a^2 = - \tfrac{1}{2} V_M^{''}(R_0)/(1 + V_M(R_0))$, so $D_e = 0.173$ and $a = 0.735$ in atomic units.
The vibrational energy levels in the Morse potential are exact, truncating at $O(v+\tfrac{1}{2})^2$,
\begin{equation*}
E_v\,=\, (v+\tfrac{1}{2})\,\w_0 - \frac{\w_0}{4 D_e}\,(v+\tfrac{1}{2})^2\,\w_0 \ ,
\end{equation*}
where the fundamental frequency is $\w_0 = a \sqrt{2D_e/\m} = 0.020$. The sign of the coefficient 
$\w_0/4D_e = 0.029$ ensures that the spacing between vibrational energy levels reduces as
$v$ increases, unlike a SHO. This parameter is equivalent to the $x_0 = 0.033$ of (\ref{f3}) calculated 
from purely anharmonic terms in the full numerical potential $V_M(R)$. This shows excellent agreement 
and gives further confidence that our analysis is providing a good characterisation of the
rovibrational spectrum at the required precision for computing SME corrections.} is also 
shown in Fig.~\ref{FigVM}.

We can also check the asymptotic behaviour of the momentum expectation values analytically. 
For both large and small $R$, where the molecule becomes effectively an atom, spherical symmetry is
restored so all the expectation values in (\ref{ap18}) and (\ref{ap19}) are equal, with
$\langle p^a p^a\rangle = \langle p^2\rangle/3 = K/3$.  This is confirmed numerically in Fig.~\ref{Figmomenta}
where we observe all the $\langle p^a p^a\rangle$ becoming equal and tending to $1/3$ for 
$R\rta \infty$ and $4/3$ for $R\rta 0$. 

\vskip0.5cm

\begin{figure}[h!]
\centering{{\includegraphics[scale=0.57]{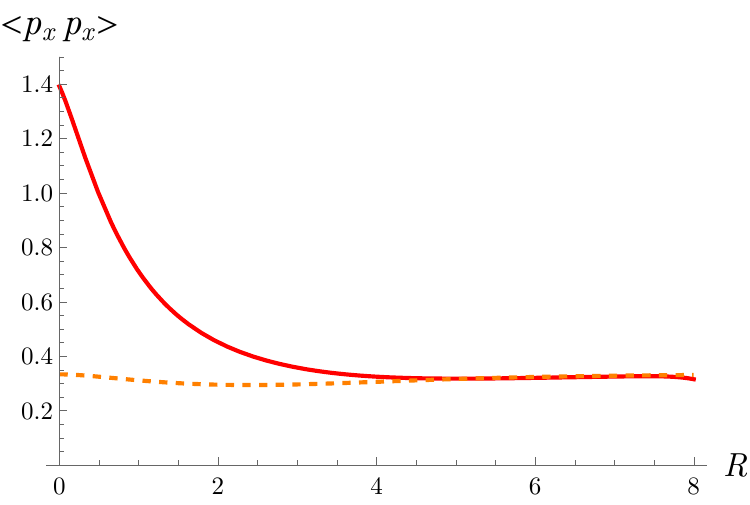}} \hskip0.5cm
{\includegraphics[scale=0.57]{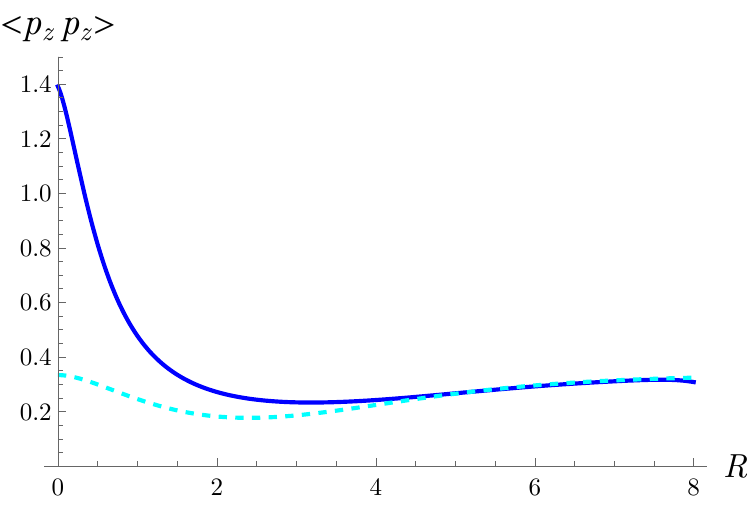}} }
\caption{The momentum expectation values from (\ref{ap18}), (\ref{ap19}) and (\ref{ap21})
as a function of the bond length $R$.
The left hand figure shows $\langle p^x\,p^x\rangle$ evaluated 
with $\gamma(R)$ and (dashed curve) fixed equal to 1.  The right hand figure is the same 
for $\langle p^z\,p^z\rangle$.}
\label{Figmomenta}
\end{figure}

\vskip0.2cm

\begin{table}[h!]
\centering
\begin{tabular} {c  l rl r l r l c l c }
%{l l l l l l l l l l l l l} %Add l for each column
\\
\hline\hline\\
 \multicolumn{1}{c} {\raisebox{1.5ex} {Derivative} } &
  \multicolumn{1}{c} {\raisebox{1.5ex} {  }  }  &
\multicolumn{1}{c} {\raisebox{1.5ex} {0}  }  &
   \multicolumn{1}{c} {\raisebox{1.5ex} {  }  }  &
\multicolumn{1}{c} {\raisebox{1.5ex} {1 } } &
   \multicolumn{1}{c} {\raisebox{1.5ex} {  }  } &
\multicolumn{1}{c} {\raisebox{1.5ex} { 2}  } &
  \multicolumn{1}{c} {\raisebox{1.5ex} {  }  }  &
\multicolumn{1}{c} {\raisebox{1.5ex} {3 }  }  &
   \multicolumn{1}{c} {\raisebox{1.5ex} {  }  }  &
\multicolumn{1}{c} {\raisebox{1.5ex} {4 }  }  
\\
\hline
\\
$V_M$ &&-1.173 &&0~~~ &&0.187 &&-0.492 &&1.237 \\
$\langle p^x\,p^x\rangle$ &&0.451 &&-0.146 &&0.146 &&-0.162  &&0.229 \\
$\langle p^z\,p^z\rangle$ &&0.271 &&-0.083 &&0.133 &&-0.188 &&0.349 \\
$\gamma$&&1.238 &&-0.203 &&0.148 &&-0.113&&0.102 \\
\\
\hline
\\
$\tr\,\langle p^a\,p^b\rangle$ &&1.173   &&-0.375  &&0.424 &&-0.512 &&0.808  \\
$\tr_{Y}\langle p^a\,p^b\rangle$ &&0.360   &&-0.126  &&0.026 &&\,\,0.052 &&-0.239 \\ 
\\
\hline\hline
\end{tabular}
\caption{The values and first few derivatives of the inter-nucleon potential $V_M(R)$, the momentum 
expectation values $\langle p^x\,p^x\rangle$ and $\langle p^z\,p^z\rangle$, and the interpolating
function $\gamma(R)$ used in their construction, all evaluated at the minimum $R_0=2.003$ of $V_M(R)$.
All numbers are expressed in atomic units, with energies in terms of the reduced Rydberg constant
$\hat{R}_H$ and lengths in terms of the reduced Bohr radius $\hat{a}_0$.  
%We also include here the combinations 
%$\tr \,\langle p^a\,p^b\rangle
%\,=\, \langle p^x\,p^x\rangle + \langle p^y\,p^y\rangle + \langle p^z\,p^z\rangle$
%and $\tr_Y \langle p^a\,p^b\rangle
%\,=\, \langle p^x\,p^x\rangle + \langle p^y\,p^y\rangle - 2\langle p^z\,p^z\rangle$
%which appear as coefficients in the SME potential $V_{\rm SME}^{e}$. 
%Recall cylindrical symmetry
%ensures that $\langle p^x\,p^x\rangle \,=\, \langle p^y\,p^y\rangle$.
}
\label{Table1}
\end{table}

\vskip0.3cm

Finally, for the analysis of the rovibrational levels in sections \ref{sect 5} and \ref{sect 6}, we need explicit
values for the first few derivatives of the energy and momentum expectation values at the minimum
$R_0$ of the improved potential $V_M(R) = E_e(R)$. These are evaluated numerically from the 
functions plotted above, and the required values are shown in Table \ref{Table1}, which may be compared
with \cite{Muller:2004tc}.
Recall that cylindrical symmetry ensures that $\langle p^x\,p^x\rangle \,=\, \langle p^y\,p^y\rangle$.
We also include here the combinations 
$\tr \,\langle p^a\,p^b\rangle
\,=\, \langle p^x\,p^x\rangle + \langle p^y\,p^y\rangle + \langle p^z\,p^z\rangle$
and $\tr_Y \langle p^a\,p^b\rangle
\,=\, \langle p^x\,p^x\rangle + \langle p^y\,p^y\rangle - 2\langle p^z\,p^z\rangle$
which appear as coefficients in the SME potential $V_{\rm SME}^{e}$.

\newpage  

\section{Rovibrational energy levels - perturbation method}\label{appendix B}

In this appendix, we present a systematic perurbative method to evaluate the rovibrational energy levels 
incorporating the Lorentz and \textsf{CPT} violating corrections. This provides an important consistency
check on the results obtained in section \ref{sect 4}. 

The idea is to expand all the contributions to the potential in the nucleon Schr\"odinger equation (\ref{c6})
about the minimum $R_0$ of the inter-nucleon potential $V_M(R)$, then evaluate the energy corrections
as perturbations about the leading SHO approximation.  As we see, this turns out to give a systematic
perturbative expansion in the small parameter $\l = 1/\mu\, \w_0 R_0^2\,$ identified previously.
We therefore write,
\begin{align}
&V_M(R) \,+\ V_N(R) \,+\, V_{\rm SME}^{e}(R)  \nonumber \\
&~~= \, V_M(R_0) \,+\, V_N(R_0) \,+\, V_{\rm SME}^{e}(R_0) \,+\, \half \mu \w_0^2 x^2 \,+\, 
\d V_M(x) \,+\, \d V_N(x) \,+\,\d V_{\rm SME}^{e}(x) \ ,
\label{apb1}
\end{align}
where here $x = R-R_0$ and $V_N(R) = N(N+1)/2\mu R^2$, so $V_N(R_0) = \tfrac{1}{2}\l \w_0 N(N+1)$.
Then,
\begin{align}
\d V_M \,&=\, \frac{1}{6} V_M^{'''}\, x^3  \,+\, \frac{1}{24} V_M^{(4)}\, x^4 \,+\, \ldots   \nonumber \\
\d V_N \,&=\, \l \w_0 N(N+1) \Big( - \frac{x}{R_0} \,+\, \frac{3}{2} \Big(\frac{x}{R_0}\Big)^2
\,-\, 2 \Big(\frac{x}{R_0}\Big)^3  \,+\, \ldots \Big)\nonumber \\
\d V_{\rm SME}^{e} \,&=\, V_{\rm SME}^{e\,'}\, x \,+\, \frac{1}{2} V_{\rm SME}^{e\,''} \,x^2 \,
+\, \frac{1}{6} V_{\rm SME}^{e\,'''}\, x^3 \,+\, \ldots
\label{apb2}
\end{align}

We start from the SHO with potential $\thalf V_M^{''}x^2 \,=\,\thalf \mu\,\w_0^2 \,x^2$ 
and construct the usual states labelled by integers $|v\rangle$ with energy eigenvalues 
$E_v^{(0)} = (v +\thalf)\w_0$. We will need the expressions for the energy levels up to 3rd order
in perturbation theory, {\it viz}.
\begin{equation}
E_v \,=\, E_v^{(0)}  \,+\, E_v^{(1)}  \,+\, E_v^{(2)}  \,+\, E_v^{(3)}    \ ,
\label{apb3}
\end{equation}
with, for a perturbation $\d V$ in the potential,
\begin{align}
E_v^{(1)} \,&=\, \langle v|\,\d V\,|v\rangle  \ ,   \nonumber \\[10pt]
E_v^{(2)} \,&=\,  \sum_{k\neq v} \,\frac{1}{(E_v ^{(0)} - E_k ^{(0)}) }\,\,
\langle v|\,\d V\,|k\rangle  \,\, \langle k|\,\d V\,|v\rangle  \ ,   \nonumber \\[10pt]
E_v^{(3)} \,&=\,\sum_{k\neq v} \sum_{\ell\neq v} \,\frac{1}{(E_v ^{(0)} - E_k ^{(0)})(E_v ^{(0)}-E_\ell ^{(0)}) }\,\, 
\langle v|\,\d V\,|k\rangle \,\, 
\langle k|\,\d V\,|\ell\rangle\, \,  \langle \ell|\,\d V\,|v\rangle  \nonumber \\
&~~~~ - \, \sum_{k\neq v} \,\frac{1}{(E_v ^{(0)} - E_k ^{(0)})^2}\,\,|\langle v|\,\d V\,|k\rangle|^2  \,\,
\langle v|\,\d V\,|v\rangle \,~\ . 
\label{apb4}
\end{align}
The states here are all understood to be the unperturbed SHO states;  clearly the notation with states
carrying a label $|v\rangle^{(0)}$ throughout is too cumbersome. 
The perturbation $\d V$  in our case is the sum of the anharmonic potential, angular momentum and 
SME terms in (\ref{b2}), $\d V \,=\, \d V_M \,+\, \d V_N \,+\,\d V_{\rm SME}^e$.

Now at first order, only even powers of $x$ contribute to the expectation values, so we have,
\begin{equation}
E_v^{(1)} \,=\, \Big( \frac{3}{2} \l \,\w_0 \,N(N+1)/R_0^2 \,+\, \frac{1}{2} \, V_{\rm SME}^{e\,''}\,\Big) \,
\langle v|\,x^2\,|v\rangle  \,   \,+\, \,
\frac{1}{24} \, \big(V_M^{(4)} \,+\, V_{\rm SME}^{e\,(4)}\big)\, \langle v|\,x^4\,|v\rangle \ .
\label{apb5}
\end{equation}
The expectation values here and throughout this section are evaluated using elementary methods,
expanding the powers of $x$ in terms of raising and lowering operators using,
\begin{equation}
\frac{1}{R_0} \,x \,=\, \sqrt{\frac{\l}{2}}\,\,(a \,+\, a^\dagger)
~~~~~~\textrm{with}~~~~~~
a^\dagger |v\rangle \,=\, \sqrt{v+1}\,\,|v+1\rangle\ ,
~~~~a |v\rangle \,=\, \sqrt{v}\,\,|v-1\rangle  \ . \\[5pt]
\label{apb6}
\end{equation}
Then, evaluating 
\begin{equation}
\frac{1}{R_0^2} \langle v|\,x^2\,|v\rangle \,=\, (v + \thalf)\,\l  \ ,~~~~~~~\textrm{and}~~~~~~
\frac{1}{R_0^4}\, \langle v|\,x^4\,|v\rangle \,=\,  \frac{3}{2} \Big(\,(v +\thalf)^2 \,+\, \tfrac{1}{4} \Big) \,\l^2 \ ,
\label{apb7}
\end{equation}
and dropping the constant term, which adds negligibly to $V_M(R_0)$, $V_{\rm SME}^{e}(R_0)$  
and the dissociation energy, we find
\begin{align}
E_v^{(1)} \,&=\,  \frac{3}{2} \, (v + \thalf)\, N(N+1) \, \l^2 \, \w_0~+~
\frac{1}{2} (v + \thalf)\, \frac{1}{V_M^{''}} \,\, V_{\rm SME}^{e\,''} \,\, \w_0  \nonumber \\
&~~~~ +\, \frac{1}{16} \,(v +\thalf)^2\,\, \frac{1}{V_M^{''}}\,R_0^2 \,\big(V_M^{(4)}\, +\,V_{\rm SME}^{e\,(4)} \,\big)
\,\,\l\, \w_0   \ .
\label{apb8}
\end{align}
We identify these as contributions to the coefficients $\a_0$, $\, \d_{SME}^e\,$ and $x_0$, $x_{\rm SME}^e$ 
respectively in section \ref{sect 5}.

Next, consider the perturbations at 2nd order.  Organising by the total number of factors of $x$ occurring
(recalling that each power of $x$ carries an associated $\sqrt{\l}$), we first find contributions in (\ref{apb4}) 
where both factors of $\d V$ are proportional to $x$; then where both $\d V \sim x^2$, together with
a mixed term with one factor of $\d V \sim x$ and the other $\d V \sim x^3$; and finally, terms with
$\d V \sim x,\, \d V \sim x^5$, $\,\d V\sim x^2, \, \d V\sim x^4$ and both $\d V \sim x^3$,
the latter three contributing leading $O(\l)$ terms to $x_0$ and $x_{\rm SME}^e$.

Taking these in turn, we first evaluate the contribution to $E_{v}^{(2)}$ from taking the perturbation
$\d V = \Big(- \l\,\w_0\,N(N+1) \,+\, R_0\,V_{\rm SME}^{e\,'}\Big) x/R_0$.  The required expectation value
is 
\begin{equation}
\sum_{k\neq v}\,   \frac{1}{(E_v ^{(0)} - E_k ^{(0)}) }\, \frac{1}{R_0^2}\, \langle v|\,x\,|k\rangle \,\,\langle k|\,x\,|v\rangle 
\,=\, -\frac{1}{2}\,\frac{1}{\big(R_0^2\,V_M^{''}\big)} \ ,
\label{apb9}
\end{equation}
with the sum over $k = v \pm 1$.  Notice that this is independent of the vibrational quantum number $v$.
The corresponding contribution to $E_{v}^{(2)}$ is therefore
\begin{equation}
E_{v;\,[1,1]}^{(2)}\,=\, - \frac{1}{2} \,\l^3 \,(N(N+1))^2\,\w_0  ~+~ 
\l \, N(N+1)\, \frac{1}{\big(R_0\,V_M^{''}\big)}\,V_{\rm SME}^{e\,'} \,\,\w_0   \ ,
\label{apb10}
\end{equation}
which we identify with terms in $D_0$ and $B_{\rm SME}^e$ respectively.

A similar calculation shows
\begin{equation}
\sum_{k\neq v}\,   \frac{1}{(E_v ^{(0)} - E_k ^{(0)}) }\, \frac{1}{R_0^4}\, \langle v|\,x\,|k\rangle \,\,\langle k|\,x^3\,|v\rangle 
\,=\, -\frac{3}{2}\,\l \,\frac{1}{\big(R_0^2\,V_M^{''}\big)} \, (v+\thalf)  \ ,
\label{apb11}
\end{equation}
and identifying the relevant perturbations $\d V$ from (\ref{apb2}) we find
\begin{align}
E_{v;\,[1,3]}^{(2)} \,&=\, -\frac{1}{2}\,(v+\thalf) \,\frac{V_M^{'''}}{\big(V_M^{''}\bigr)^2} \, V_{\rm SME}^{e\,'} \, \,\w_0  ~+~
\frac{1}{2}\,\l^2\,(v+\thalf) \,N(N+1)\, \frac{R_0\, V_M^{'''}}{V_M^{''} }\, \w_0 \nonumber \\
&~~~~~~~ +\, \l^2\, (v+\thalf) \,N(N+1)\, \Big( \frac{1}{2} \frac{R_0}{V_M^{''}}\,\, V_{\rm SME}^{e\,'''} 
\,+\, 6 \frac{1}{\big(R_0\,V_M^{''}\big)}\, V_{\rm SME}^{e\,'} \,\Big)\,\,\w_0\ .
\label{apb12}
\end{align}
These are contributions to $\d_{\rm SME}^e$,  $\a_0$ and $\a_{\rm SME}^e$ respectively.

The next contribution at 2nd order comes from terms where both $\d V\sim x^2$, so here we need to
calculate
\begin{equation}
\sum_{k\neq v}\,   \frac{1}{(E_v ^{(0)} - E_k ^{(0)}) }\, \frac{1}{R_0^4}\, \langle v|\,x^2\,|k\rangle \,\,
\langle k|\,x^2\,|v\rangle 
\,=\, -\frac{1}{2}\,\l \,\frac{1}{\big(R_0^2\,V_M^{''}\big)} \, (v+\thalf)  \ ,
\label{apb13}
\end{equation}
giving the energy
\begin{equation}
 E_{v;\,[2,2]}^{(2)} \,=\, -\frac{3}{4} \,\l^2\, (v+\thalf) \,N(N+1)\, \frac{1}{V_M^{''}}\,V_{\rm SME}^{e\,''}\,\,\w_0 \ ,
\label{apb14}
\end{equation}
which adds to $\a_{\rm SME}^e$.  We have discarded the
$\l^4\, (v+\thalf)\,\big(N(N+1)\big)^2\w_0$ term
here as it is not of the type for which we have kept the coefficients
in (\ref{e8}).

Moving on to the 2nd order contributions with a total of 6 powers of $x$, we find they involve factors
of $(v +\thalf)^2$, so we are only concerned here with their effect on $x_0$ and $x_{\rm SME}^e$.
In turn, these are:
\begin{equation}
\sum_{k\neq v}\,   \frac{1}{(E_v ^{(0)} - E_k ^{(0)}) }\, \frac{1}{R_0^6}\, \langle v|\,x^3\,|k\rangle \,\,\langle k|\,x^3\,|v\rangle 
\,=\, -\frac{15}{4}\,\l \,\w_0\, \big[(v+\thalf)^2 \,+\, \tfrac{7}{60}\big]\, \frac{1}{\big(R_0^2\,V_M^{''}\big)^2} \ ,
\label{apb15}
\end{equation}
from which,
\begin{equation}
E_{v;\,[3,3]}^{(2)} \,=\, - \frac{5}{48} \, \l\,\w_0\,\big[(v+\thalf)^2 \,+\, \tfrac{7}{60}\big]\,
\frac{1}{V_M^{''}}\,R_0^2\, \big( (V_M^{'''})^2 \,+\, 2   V_M^{'''}\, V_{\rm SME}^{e\, '''}\big) \ ,
\label{apb16}
\end{equation}
contributing to $x_0$ and $x_{\rm SME}^e$;
\begin{equation}
\sum_{k\neq v}\,   \frac{1}{(E_v ^{(0)} - E_k ^{(0)}) }\, \frac{1}{R_0^6}\, \langle v|\,x^2\,|k\rangle \,\,\langle k|\,x^4\,|v\rangle 
\,=\, -\frac{3}{2}\,\l \,\w_0\, \big[(v+\thalf)^2 \,+\, \tfrac{1}{4}\big]\, \frac{1}{\big(R_0^2\,V_M^{''}\big)^2} \ ,
\label{apb17}
\end{equation}
giving
\begin{equation}
E_{v;\,[2,4]}^{(2)} \,=\, - \frac{1}{16} \, \l\,\w_0\,\big[(v+\thalf)^2 \,+\, \tfrac{1}{4}\big]\,
\frac{1}{\big(V_M^{''}\big)^2}\,R_0^2\,  V_M^{(4)} \, V_{\rm SME}^{e\, ''}  \ ,
\label{apb18}
\end{equation}
which contributes only to $x_{\rm SME}^e$; and
\begin{equation}
\sum_{k\neq v}\,   \frac{1}{(E_v ^{(0)} - E_k ^{(0)}) }\, \frac{1}{R_0^6}\, \langle v|\,x\,|k\rangle \,\,\langle k|\,x^5\,|v\rangle 
\,=\, -\frac{15}{4}\,\l \,\w_0\, \big[(v+\thalf)^2 \,+\, \tfrac{1}{4}\big]\, \frac{1}{\big(R_0^2\,V_M^{''}\big)^2} \ ,
\label{apb19}
\end{equation}
giving
\begin{equation}
E_{v;\,[1,5]}^{(2)} \,=\, - \frac{1}{16} \, \l\,\w_0\,\big[(v+\thalf)^2 \,+\, \tfrac{1}{4}\big]\,
\frac{1}{\big(V_M^{''}\big)^2}\,R_0^2\,V_M^{(5)}\, V_{\rm SME}^{e\, '}  \ ,
\label{apb20}
\end{equation}
which again only contributes to $x_{\rm SME}$.   Comparing with (\ref{e9}) and (\ref{e19}), we see that we have 
now reproduced exactly all the terms found in the effective potential method up to 2nd order in the
perturbations.

So far, in (\ref{apb7}) and (\ref{apb15}), (\ref{apb17}), (\ref{apb19}), we have just discussed the $O(v+\thalf)^2$
contributions, keeping only the terms independent of the angular momentum since we dropthese at 
$O\big((v+\thalf)^2 N(N+1)\big)$.  However, the constant factors {\it do} contribute to $B_0$, $B_{\rm SME}^e$
and $D_0$, though not $D_{\rm SME}^e$ since this only receives contributions at 3rd order (see below).
Inspecting these terms, we see that they give {\it sub-leading} contributions, of $O(\l^3)$ in $B_0$
and $B_{\rm SME}^e$ and $O(\l^5)$ in $D_0$, picking up all the terms of 2nd order, but only 2nd order, 
in the perturbations.  This confirms that in (\ref{e9}),
$B_0 = \frac{1}{2}\l \,+\, O(\l^3)$ is a power series in $\l^2$ with leading term of $O(\l)$,
with an analogous result for $B_{\rm SME}^e$ and $D_0$.

\vskip0.3cm

Evidently we can continue straightforwardly to as many orders as desired, though of course evaluating
the products of expectation values becomes increasingly laborious.  It is interesting however to display
just one term at 3rd order, which gives the leading term to the coefficient $D_{\rm SME}^e$.
In this case we need two factors of $\d V\sim x$ and one $\d V\sim x^2$ and from (\ref{apb4}) the
required calculation is
\begin{align}
&\sum_{k\neq v} \sum_{\ell\neq v} \,\frac{1}{(E_v ^{(0)} - E_k ^{(0)})(E_v ^{(0)}-E_\ell ^{(0)}) }\,\, \frac{1}{R_0^4}\,\Big[
\langle v|\,x\,|k\rangle \,\, \langle k|\,x^2\,|\ell\rangle\, \,
  \langle \ell|\,x\,|v\rangle \nonumber \\
&~~~~~~~~~~~~~~~~~~~~~~~~~~~~~~~~~~~~~~~~~~~~~~~~~~~~~~~~~~~~~~~~~~
\,+\,2\, \langle v|\,x\,|k\rangle \,\, \langle k|\,x\,|\ell\rangle\, \,  \langle \ell|\,x^2\,|v\rangle \,\Big]
\nonumber \\
&~~  - \sum_{k\neq v}\,   \frac{1}{(E_v ^{(0)} - E_k ^{(0)})^2 }\,\,\frac{1}{R_0^4}\,
\langle v|\,x\,|k\rangle \,\, \langle k|\,x\,|v\rangle\, \,  \langle v|\,x^2\,|v\rangle  \nonumber \\[5pt] 
&=\,\, \frac{1}{4}\,\frac{1}{\big(R_0^2 V_M^{''}\big)^2}\,\,
\Big[\big(2(v+\thalf)^2 + \tfrac{5}{2}\big)\,+\, 2\big( (v + \thalf)^2
  + \tfrac{3}{4}\big) \,-\, 4(v+\thalf)^2 \Big] 
\nonumber \\
&= \,\, \frac{1}{\big(R_0^2 V_M^{''}\big)^2}  \ .
\label{apb21}
\end{align}
Remarkably, the terms of $O(v+\thalf)^2$ cancel, leaving a residual $v$-independent term, as required
to contribute to the $D_{\rm SME}^e$ coefficient.  Then, since we can select terms with both 
$\d V_{\rm SME}^e\sim x^2, \,\d V_N\sim x, \,\d V_N\sim x$ and $\d V_{\rm SME}^e\sim x, \,\d V_N\sim x,
\,\d V_N\sim x^2$, we identify the required contribution to $E_v^{(3)}$:
\begin{equation}
E_v^{(3)} \,=\, \l^3\, \big((N(N+1)\big)^2\,\,
\bigg[\,\frac{1}{2} \frac{1}{V_M^{''}}\, V_{\rm SME}^{e\,''}  \,-\, 
3  \frac{1}{\big(R_0\,V_M^{''}\big)}\, V_{\rm SME}^{e\,'}  \,\bigg]\,\w_0 ~~~\ldots  
\label{apb22}
\end{equation}
This is to be compared with (\ref{e18}) calculated with the effective potential method.

A careful comparison of these results with those derived by the effective potential method in 
section \ref{sect 5} shows that apart from (\ref{apb22}) we have yet to identify the terms in 
(\ref{e9}) and (\ref{e18}),(\ref{e19}) involving more than two perturbative factors,
frequently involving higher derivatives of the potential $V_M(R)$.
To compare the origin of these terms in the effective potential and perturbation theory methods, and 
to verify that the terms quoted in section \ref{sect 5} for the rovibrational energies are {\it complete}
at the quoted order in $\l$, we need to analyse more closely the systematics of the expansion in $\l$
in this expectation value approach.

First, notice that in the perturbation series in (\ref{apb4}), each power of $x$ in the perturbations
$\d V$ brings a factor of $\sqrt{\l}$ from its expression (\ref{apb6}) in raising and lowering operators.
Then from the energy denominators, each further order in $\d V$ brings with it an extra factor
of $1/\w_0 = 1/ (R_0^2\, V_M^{''} \,\l)$.  

We can therefore count the relevant orders by inspection. For example, consider the $D_0$ coefficient.
This arises first at 2nd order from (\ref{b9}) with both $\d V_N \sim x$, so the expectation value itself is
$O(\l/\w_0$ and since $\d V_N$ itself is of $O(\l \w_0$, the corresponding energy is $O(\l^3 \w_0)$.
Remembering to extract a factor $w_0$ to leave the coefficient $D_0$ dimensionless, 
we deduce immediately that $D_0 \sim \l^3$, as given in (\ref{b9}.

The next contribution would come from including the $O(x^3)$ term in the expansion of one
of the $\d V_N$ factors. But then the expectation value would be $O(\l^2/\w_0$ and the contribution 
to $D_0$ would be $O(\l^4$. We see therefore that as already indicated, the coefficient $D_0$
quoted in (\ref{b9}) is the leading term in a perturbative series in the small parameter $\l$.
The same is true of all the rovibrational energy coefficients.

We can now use this power counting to understand the origin of the factors involving higher derivatives 
of $V_M(R)$ and why they must be included.   A good example is $\a_0$. As we have seen above, this arises
already in 1st order with $\d V_N \sim \l \w_0 x^2$.  The expectation value $\langle x^2\langle$ is $O(\l)$,
so we find the $E_v^{(1)}$ contribution is $O(\l^2 \w_0$, giving $\a_0 \sim \l^2$.
However, from (\ref{b9}) we know that this is not the complete result for $\a_0$ which has another
contribution of the same $O(\l^2)$ but with a factor depending on $V_M^{'''}$.
This arises at 2nd order, with perturbations $\d V_N \sim \l \,\w_0 x$ and $\d V_M \sim V_M^{'''} \,x^3$.
Following the power counting rules, this $E_{v}^{(2)}$ contribution is $O(\l^2 \w_0 V_M^{'''}/V_M^{''})$,
giving the complete $O(\l^2)$ result for $\a_0$ found in (\ref{b9}).\footnote{In keeping track of these orders,
we always express the energies with a single overall factor of $\w_0$, with all other occurrences of $\w_0$
being traded for derivatives $V_M^{''}$ using the relation $\w_0 = R_0^2 V_M^{''} \l$.
This isolates the correct order in $\l$ leaving residual factors involving ratios of derivatives of $V_M$
as in (\ref{b9}) for $\a_0$; numerically, we have found that these ratios are of $O(1)$.
So this method of arranging the series correctly groups contributions of the same magnitude.}

What this example shows us is that in general we can find another contribution to a particular
rovibrational coefficient which is of the same order in $\l$ by going to a higher order in the SHO perturbation
series (\ref{apb4}) and including a correspondingly higher derivative term in the expansion of $\d V_M$
{\it if} (and only if) at the same time we can {\it reduce} the power of $x$ from the expansion of the 
perturbation $\d V_N$ or $\d V_{\rm SME}$ which specifies that  coefficient. This process can obviously 
not be iterated indefinitely and this is why the number of terms in a rovibrational coefficient
at a given order in $\l$ is limited.

These considerations allow us to identify the perturbative origin of {\it all} the terms in the
rovibrational coefficients quoted in section \ref{sect 5} and to verify that they are indeed 
complete.\footnote{To verify this on the most complicated example, consider the expression for 
$\a_{\rm SME}^e$ in (\ref{e18}).  Label the terms in the order given there as (1) to (7).
By inspection, their origin is as follows. At 2nd order,
~(1)~$\d V_N \sim x$, $\d V_{\rm SME}^e \sim x^3$ ;
~(2)~$\d V_N\sim x^2$, $\d V_{\rm SME}^e \sim x^2$ ;
~(4)~$\d V_N\sim x^3$, $\d V_{\rm SME}^e \sim x$.
Then at 3rd order, we find the terms involving the expansion of $\d V_M$:
~(3)~$\d V_N \sim x$, $\d V_{\rm SME}^e \sim x^2$, $\d V_M \sim x^3$ ;
~(5)~$\d V_N \sim x^2$, $\d V_{\rm SME}^e \sim x$, $\d V_M \sim x^3$ ; 
~(7)~$\d V_N \sim x$, $\d V_{\rm SME}^e \sim x$, $\d V_M \sim x^4$.
Finally, there is one possible further term at 4th order:
~(6)~$\d V_N \sim x$, $\d V_{\rm SME}^e \sim x$, $\d V_M \sim x^3$, $\d V_M \sim x^3$.
At this point we cannot iterate further by introducing more orders in $\d V_M$ because we cannot
reduce the powers of $x$ in $\d V_N$ and $\d V_{\rm SME}^e$ any more.  Any further terms 
are of higher order in $\l$. We have therefore 
found the {\it complete} set of terms in $\a_{\rm SME}^e$ at its leading order in $\l$, {\it i.e.} $O(\l^2)$. 
A similar exercise readily identifies the origin of all the terms in the expression (\ref{e19}) for $x_{\rm SME}^e$. }
Notice also that the effective potential method, where we analyse the anharmonic oscillator
around the minimum of the potential {\it including} the perturbations, is in some ways more efficient
in identifying terms that otherwise require high orders in the systematic perturbation method.

}

\end{document}